\documentclass[aps,prb,twocolumn,showpacs,amsmath]{revtex4}
\usepackage{latexsym}
\usepackage{amssymb}
\usepackage{graphicx}
\usepackage{bm}
\usepackage[english]{babel}
\usepackage[latin1]{inputenc}
\usepackage{amsmath}
\usepackage{amsfonts}

\newcommand{\bq}{\begin{equation}}
\newcommand{\eq}{\end{equation}}
\newcommand{\bn}{\begin{eqnarray}}
\newcommand{\en}{\end{eqnarray}}

\begin{document}

\title{Full Counting Statistics of Electron Tunneling through Coherently Coupled Quantum Dots: Exchange Interaction Effect on 
Shot Noise}

\author{Bo Xiong}
\affiliation{School of Agriculture and Biology, Shanghai Jiao Tong University,
800 Dongchuan Road, Shanghai 200240, China}

\author{Guo-Hui Ding}
\affiliation{Department of Physics, Shanghai Jiao Tong University,
800 Dongchuan Road, Shanghai 200240, China}

\author{Bing Dong}
\thanks{Author to whom correspondence should be addressed. Email:bdong@sjtu.edu.cn.}
\affiliation{Department of Physics, Shanghai Jiao Tong University,
800 Dongchuan Road, Shanghai 200240, China}

\begin{abstract}
We investigate the zero-frequency shot noise of electronic tunneling through a single quantum dot (SQD) and coherently coupled 
quantum dots (CQD) taking into account the Coulomb interaction. Within Hartree-Fock approximation, the analytical expressions of 
current and zero-frequency shot noise are self-consistently derived in the framework of full counting statistics for the both 
systems. We demonstrate that the correction term of zero-frequency shot noise induced by the intradot Coulomb interaction is 
almost negligible compared to the noninteracting shot noise in a SQD, while in a CQD the interplay of the interdot coherence 
fluctuations and strong interdot Coulomb interaction can induce a super-Poissonian noise even in the symmetric case.

\end{abstract}



\maketitle

\section{INTRODUCTION}

The dynamics of electron transport through the quantum dot (QD) system is one of the most interesting topics in mesoscopic 
physics, which has motivated many experimental investigations, such as ones on the structures of barriers with localized 
states,\cite{Safonov,chen1} the single quantum dot(SQD)\cite{Onac,Ubbelohde,Cottet,zhang,Zarchin} and the double or multiple 
quantum dots,\cite{Barthold,Ubbelohde1,Muller,Robledo,Kieblich1,McClure} and also a lot of 
theoretical research.\cite{Blanter,Thielmann,Sukhorukov,Gustavsson,Gattobigio,Kieblich3,Belzig,Djuric,Aghassi,Dong1,Dong2,Lopez} 
In particular, the shot noise, resulting from charge discreteness and stochastic transfer process, serves as an important 
indicator for the dynamic property in the mesoscopic structures,\cite{Blanter} as well as a revelation of a lot of other 
information, such as the charge unit\cite{Lefloch,Kane} and entanglement\cite{Burkard,Lesovik} of the transferred charges. The 
enhancement of shot noise is an important signature for the non-Poisson process. The classical Poisson process requires shot 
noise $S$ follows the expression $S=2 e I$, where $e$ refers to the unit charge and $I$ the current, while in quantum transport 
the charge distribution is proved to follow the binomial statistics.\cite{S. Levitov} It has been reported that the electronic 
tunneling through a single quantum dot with multiple energy levels and a coupled quantum dots (CQD) system turned outs to be 
super-Poisson ($S > 2 e I$).\cite{zhang,Zarchin,Barthold,Kieblich1} For the system of CQD, the interdot coherence and Coulomb 
interaction are essential factors to determine the shot noise characteristics of this system, thereby arousing great research 
interests\cite{Barthold,Kieblich1,Djuric,Aghassi,Dong1,Dong2} and deserving more research investigations.

Over the past decades, many theoretical attempts have been devoted to understanding of the super-Poisson process in the QD 
system, especially the effects of Coulomb blockade in coherently coupled quantum dots system on the enhancement of zero-frequency 
shot noise. It is worthwhile to point out that when only the interdot coherent coupling\cite{Kieblich2,Aguado} or the Coulomb 
blockade\cite{Hershfield1,Bagrets,Thielmann1} effect is taken into account, the noise-to-current ratio (Fano factor) is found to 
be suppressed.
However, the joint effect of the coherence and the Coulomb interaction has been reported to enhance the shot noise. Aghassi {\it 
et al.}\cite{Aghassi} demonstrated the appearance of super-Poissonian noise for symmetric configuration of a CQD taking account 
of both intra- and inter-dot Coulomb repulsions if the vertical energy gap is relatively smaller than the sequential energy gap 
and the applied bias voltage is below the sequential threshold. Nevertheless for the strongly broken symmetric case, they 
reported the observation of both super-Poissonian noise and negative differential conductance in the region with the bias above 
the sequential tunneling threshold. Kie{\ss}lich {\it et al.}\cite{Kieblich1} attributed the noise enhancement of CQD to the 
interplay of strong Coulomb repulsion and quantum coherence, and showed the high sensitivity of such mechanism to the decoherence 
caused by electron-phone scattering. One of the authors, Dong,\cite{Dong1,Dong2} predicted a giant Fano factor of electronic 
tunneling through a parallel CQD in the strong Coulomb blockade regime due to the interference between two transport path.  
However, all the works mentioned above employed the technique of master equation, whether modified or not, whose validity is not 
ensured in the small bias region and strong interacting system. Out of the application boundary of master equation, more 
appropriate techniques are expected. For instance,  L\'{o}pez {\it et al.}\cite{Lopez} adopted nonequilibrium Green funtions 
(NGF) with slave-boson mean-field theory to describe correlated double quantum dots system.

In this work, within the framework of NGF and Hartree-Fock (HF) decoupling approximation, we adopt the full counting statistics 
(FCS) theory \cite{Nazarov,Levitov1,Gogolin1,Gogolin2} to investigate the zero-frequency shot noise of electronic tunneling 
through a series CQD with symmetric configuration in strong Coulomb blockade regime. FCS was introduced to describe the 
probability distribution of charge transport through mesoscopic conductors and thus contained important dynamic 
information.\cite{S. Levitov} With these instruments, we study the joint effect of coherent exchange and Coulomb interactions on 
the shot noise in detail. Due to this joint effect, super-Poissonian characteristic structures in shot noise, i.e. the peak and 
plateau, are found at the small and large bias region, respectively.

The outline of this paper is as follows.  In Sec.II, we give a simplified approach to zero-frequency shot noise in single quantum 
dot in comparison with the work of Hershfield,\cite{Hershfield} who performed the derivations of shot noise expression within the 
diagrammatic technique. In this way, we confirm the validity of our theory to some extent. In Sec.III, this technique is 
generalized to study the CQD system, and numerical simulations and discussions are presented. Then the summary is given in 
Sec.IV.

\section{SINGLE QUANTUM DOT}

\subsection{Model and Approximation}

We adopt the impurity Anderson model to describe the single quantum dot system including two electrodes (source and drain 
reservoirs) and a quantum dot, specifically as
\begin{equation}
H = H_{L} + H_{R} + H_{D} + H_{T},
\end{equation}
with
\begin{equation}
H_{\eta}=\sum_{\eta k \sigma} \epsilon_{\eta k \sigma} c_{\eta k \sigma}^{\dagger} c_{\eta k \sigma},
\end{equation}
\begin{equation}
H_{D}=\sum_{\sigma} E_{0} d_{\sigma}^{\dagger} d_{\sigma} + U d_{\uparrow}^{\dagger} d_{\uparrow} d_{\downarrow}^{\dagger} 
d_{\downarrow},
\end{equation}
\begin{equation}
H_{T} = \sum_{\eta k \sigma} \left[ V_{\eta k}e^{i\lambda_{\eta}(t)/2} d_{\sigma}^{\dagger} c_{\eta k \sigma} + H.c. \right] .
\end{equation}
Here, $H_{\eta}$ ($\eta=\{L,R\}$) models the left and right leads with the noninteracting approximation, where $\epsilon_{\eta k 
\sigma}$ denotes the single-electron energy with the momentum label ($k$) and spin label ($\sigma=\{ \uparrow , \downarrow \}$) 
and $c_{\eta k \sigma}^{\dagger}$ ($c_{\eta k \sigma}$) creates (destroys) a corresponding electron in the electrode $\eta$. 
$H_{D}$ describes the quantum dot with the electron energy $E_{0}$ and the Coulomb interaction parameter $U$, where 
$d_{\sigma}^{\dagger}$ and $d_{\sigma}$ are the creation and annihilation operators for the spin-$\sigma$ electron on the dot. 
The last term $H_{T}$ models the electron tunneling between leads and the dot with the coupling parameters $V_{L k}$ and $V_{R 
k}$, which link the dot to the left electrode and the right one respectively. In the framework of FCS, the counting field with 
respect to the lead $\eta$ is introduced and labeled by $\lambda_{\eta}(t)$. Note that in this work, the symbol $-$ denotes the 
forward path on Keldysh contour and the symbol $+$ the backward path. Thus, on the forward or backward path $\lambda_{\eta}(t)$ 
is expressed as $\lambda_{\eta}(t)=\lambda_{\eta \mp}\theta(t) \theta({\cal T}-t)$ and furthermore results in $\lambda_{\eta 
-}=-\lambda_{\eta +}=\lambda_{\eta}$ in calculations of FCS. Here, ${\cal T}$ denotes the measuring time during which the 
counting fields are non-zero. Because of the configuration symmetry, we might as well set the counting fields between the right 
lead and
the dot $\lambda_{R -}=\lambda_{R +}=0$ for the convenience.\cite{Levitov1,Gogolin1,Gogolin2}

For the purpose of investigating FCS, we employ the Hartree approximation to deal with the Coulomb interaction within the quantum 
dot, i.e. write the second term of $H_{D}$ as $U \rho_{\uparrow} d_{\downarrow}^{\dagger} d_{\downarrow} + U \rho_{\downarrow} 
d_{\uparrow}^{\dagger} d_{\uparrow}$, where we define $\rho_{\uparrow}=\langle d_{\uparrow}^{\dagger} d_{\uparrow} 
\rangle_{\lambda}$ and $\rho_{\downarrow}=\langle d_{\downarrow}^{\dagger} d_{\downarrow} \rangle_{\lambda}$. It is deserved to 
emphasize that $\rho_{\sigma}^{-} \neq \rho_{\sigma}^{+}$, whose superscripts $-$ and $+$ indicate the position of time argument 
on the Keldysh contour, resulting from the fact $\lambda_{\eta -} \neq \lambda_{\eta +}$. Therefore, in this approximation the 
total $H_{D}$ is written as
\begin{eqnarray}
H_{D} &=& \sum_{\sigma} E_{0} d_{\sigma}^{\dagger} d_{\sigma} + U \rho_{\uparrow} d_{\downarrow}^{\dagger} d_{\downarrow} + U 
d_{\uparrow}^{\dagger} d_{\uparrow} \rho_{\downarrow} \nonumber\\
 &=& \sum_{\sigma} \epsilon_{0} d_{\sigma}^{\dagger} d_{\sigma},
\end{eqnarray}
where we introduce the $\lambda$-dependent noninteracting energy level $\epsilon_{0}$, which is defined as $\epsilon_{0} = 
E_{0}+U\rho$. Here, the $\lambda$-dependent electron occupation number holds the relation that 
$\rho=\rho_{\uparrow}=\rho_{\downarrow}$ in this non-magnetic system. However, note that $\epsilon_{0}^{-} \neq 
\epsilon_{0}^{+}$, from the obvious relation $\rho^{-} \neq \rho^{+}$. At the end, we point out that through this paper we use 
the natural units ($\hbar=e=k_{B}=1$).

\subsection{Theoretical Formulation}

In this subsection, we shall first carry out the calculation to obtain the generating function $\chi(\lambda)$ or the cumulant 
generating function $\ln\chi(\lambda)$. Then according to the relation between cumulants and the cumulant generating function, 
the current and the shot noise expressions are derived respectively, actually ending up in terms of some response and correlation 
functions which will be defined later.

To obtain the so-called cumulant generating function $\ln\chi(\lambda)$, which takes the expression as $\chi(\lambda)= \langle 
\textit{T}_C e^{-i \int_C H_T(t) dt} \rangle$, where $\textit{T}_{C}$ is the Keldysh contour ordering operator, the adiabatic 
method is a good choice: $\ln \chi(\lambda)= - i{\cal T} {\cal U}(\lambda_{-}, \lambda_{+})= - i {\cal T} {\cal U}(\lambda, 
-\lambda)$, where ${\cal U}(\lambda_{-}, \lambda_{+})$ is the adiabatic potential. A quite straightforward way is to take the 
functions $\lambda_{\pm}(t)$ to different constants as $\lambda_{\pm}$ first, then substitute the expression of $T_{\lambda}$ 
into the non-equilibrium Feynman-Hellmann equation, which is
${\partial {\cal U}(\lambda_{-},\lambda_{+}) \over {\partial \lambda_{-}}} = \Big\langle {\partial H_{T}(t)\over{\partial
\lambda_{-}}} \Big\rangle_\lambda$, and naturally arrive at
\bq \label{1}
{\partial {\cal U}(\lambda_{-},\lambda_{+}) \over {\partial \lambda_{L -}}} = {i \over 2}\sum_{k \sigma} \langle V_{L k} 
e^{i\lambda_{L -}/2} d_{\sigma}^{\dagger}  c_{L k \sigma} - {\rm H.c.}\rangle_{\lambda}.
\eq
For practical reasons, we introduce the contour-order mixed GFs
\begin{equation}
G_{\eta k \sigma, d \sigma}(t,t')=-i\langle{ \textit{T}_{C} c_{\eta k \sigma}(t) d_{\sigma}^{\dagger}(t')} \rangle_{\lambda},
\end{equation}
\begin{equation}
G_{d \sigma, \eta k\sigma}(t,t')=-i\langle{ \textit{T}_{C} d_{\sigma}(t) c_{\eta k\sigma}^{\dagger}(t')} \rangle_{\lambda}.
\end{equation}
Thus, we can rewrite Eq.(\ref{1}) as
\begin{widetext}
\bq \label{adiabatic}
 {\partial {\cal U}(\lambda_{-},\lambda_{+})\over{\partial \lambda_{L -}}} = {1\over{2}} \sum_{k\sigma} \left [  V_{L k} 
e^{i\lambda_{L -}/2} G_{L k \sigma, d \sigma}^{--}(t,t^{+}) - V_{L k} e^{-i\lambda_{L -}/2} G_{d \sigma, L k 
\sigma}^{--}(t,t^{+}) \right ].
\eq
where $t^{+}$ is defined as $t^{+}=t+0^{+}$. Note that
\bq
G_{\eta k \sigma, d \sigma}^{--}(t,t') = V_{Lk} \int dt_{1} \left[ g_{\eta k\sigma}^{--}(t,t_{1}) e^{-i\lambda_{\eta -}/2} 
G_{d\sigma}^{--}(t_{1},t') - g_{\eta k\sigma}^{-+}(t,t_{1}) e^{-i \lambda_{\eta +}/2} G_{d\sigma}^{+-}(t_{1},t') \right],
\eq
\bq
G_{d \sigma ,\eta k \sigma}^{--}(t,t') = V_{Lk} \int dt_{1} \left[ G_{d\sigma}^{--}(t,t_{1}) e^{i \lambda_{\eta -}/2} g_{\eta 
k\sigma}^{--}(t_{1},t') - G_{d\sigma}^{-+}(t,t_{1}) e^{i \lambda_{\eta +}/2} g_{\eta k\sigma}^{+-}(t_{1},t') \right],
\eq
where we define the QD Green function $G_{d\sigma}(t,t') = -i\langle {\textit{T}_{C}} d_{\sigma}(t) d_{\sigma}^{\dagger}(t') 
\rangle_{\lambda}$ and the bare lead GFs are $g_{\eta k \sigma}(t,t') = -i\langle {\textit{T}_{C}} c_{\eta k \sigma}(t) c_{\eta k 
\sigma}^{\dagger}(t') \rangle_{\lambda}$. Substituting the above two Dyson equations into Eq.(\ref{adiabatic}) and performing 
Fourier transformation, we obtain
\bq
 {\partial {\cal U}(\lambda_{-},\lambda_{+}) \over {\partial \lambda_{-}}} = {i \over 2} \sum_{\sigma} \int {dE \over{2 \pi}} 
\left[ G_{d\sigma}^{-+}(E) \Sigma_{L\sigma}^{+-}(E) - \Sigma_{L\sigma}^{-+}(E) G_{d\sigma}^{+-}(E) \right], \label{adiabatic2}
\eq
where with superscripts $\mu, \nu$ marking the time position on the Keldysh contour($\mu, \nu=\{-,+\}$), we define the 
self-energy
\bq
\Sigma_{\eta \sigma}^{\mu\nu}(E) = \sum_{k} V_{\eta k}^{2} e^{i(\lambda_{\eta \mu}-\lambda_{\eta \nu})/2} g_{\eta 
k\sigma}^{\mu\nu}(E).
\eq
From now on, we will omit the energy argument $E$ to save the space. Obviously, in order to obtain the adiabatic potential, we 
have to evaluate the GFs for the QD. Using the equation of motion (EOM) technique (details please refer to the Appendix A), we 
obtain
\begin{equation}\label{sqdGexpression}
\hat{G}_{d\sigma}=
\left(
\begin{array}{cc}
G_{d\sigma}^{--} & G_{d\sigma}^{-+} \\ G_{d\sigma}^{+-} & G_{d\sigma}^{++}
\end{array}
\right)
= {1 \over{\cal D}}
\left(
\begin{array}{cc}
E-E_{0}-U\rho^{+} + \sum_{\eta} i \Gamma_{\eta} (f_{\eta}-{1 \over{2}}) & i e^{i\bar{\lambda}_{L}/2}\Gamma_{L} f_{L} + i 
\Gamma_{R} f_{R} \\
 -i \Gamma_{L} e^{-i\bar{\lambda}_{L}/2} (1-f_{L}) -i \Gamma_{R} (1-f_{R}) & -(E-E_{0}-U\rho^{-}) + \sum_{\eta} i \Gamma_{\eta} 
(f_{\eta}-{1 \over{2}})
\end{array}
\right),
\end{equation}
where $\bar{\lambda}_{\eta} = \lambda_{\eta -} - \lambda_{\eta +}$. Here ${\cal D}$ is given as
\bn
{\cal D} &=& i \left[ \Gamma_{R}(f_{R}-{1 \over{2}})+ \Gamma_{L}(f_{L}-{1 \over{2}}) \right] (\epsilon_{0}^{+}-\epsilon_{0}^{-})
  + (E-\epsilon_{0}^{-})(E-\epsilon_{0}^{+}) + \Gamma^{2} \cr
  && -\Gamma_{L}\Gamma_{R}  \left[ f_{R}(1-f_{L})(1-e^{-i\bar{\lambda}_{L}/2})+f_{L}(1-f_{R})(1-e^{i\bar{\lambda}_{L}/2}) 
\right],
\en
where $f_{\eta}$ denotes the Fermi distribution function with respect to the lead $\eta$ and $\Gamma$ is defined as $\Gamma = 
(\Gamma_{L}+\Gamma_{R})/2$ with the definition of level-width function $\Gamma_{\eta}=2 \pi \sum_{k}\rho(E_{k}) V_{\eta k}^2$ and 
$\rho(E_{k})$ indicating the density of states. $\Gamma_{\eta}$ can be taken as energy-independent and hence a constant with the 
assumption of the wide band limit.
\end{widetext}
Solve the Dyson equation and then obtain the demanded expressions of $G_{d\sigma}^{-+}$ and $G_{d\sigma}^{+-}$ for the evaluation 
of Eq.(\ref{adiabatic2}). To complete the calculations for the adiabatic potential, it should be noted that the Hartree 
approximation introduces two unknown parameters $\rho^{-}$ and $\rho^{+}$ involved in the expression of ${\cal D}$ and hence in 
that of $\hat{G}_{d\sigma}$. To solve the problem, we have to find the self-consistent equations and the corresponding solution 
of GFs, which we need to derive the adiabatic potential formula. According to the definition of $G_{d\sigma}^{--}(t,t')$ and 
$G_{d\sigma}^{++}(t,t')$, we obtain the following self-consistent equations:
\begin{equation}
\rho^{-} = \int {{dE}\over{2 \pi i}} G_{d\sigma}^{--} e^{i E 0^{+}}, \label{selfconsissqd1}
\end{equation}
\begin{equation}
\rho^{+} = \int {{dE}\over{2 \pi i}} G_{d\sigma}^{++} e^{-i E 0^{+}}. \label{selfconsissqd2}
\end{equation}
 It is found that the to-be-solved parameters $\rho^{-}$ and $\rho^{+}$ also contained in the expressions for $G_{d\sigma}^{--}$ 
and $G_{d\sigma}^{++}$. Then one can do the expansion of $\rho^{-}$ and $\rho^{+}$ in the Eq.(\ref{selfconsissqd1}) and 
Eq.(\ref{selfconsissqd2}) with respect to the counting field $\lambda$ to some proper orders according to the aim of the 
calculations.
In the following, we just need to expand them to the first order with respect to $({i \over 2}\lambda)$:\cite{Dong3}
\begin{equation}
\rho^{-}=\rho^{(0)} + \rho^{-(1)}\left({i\over 2}\lambda\right) + o(i\lambda),
\end{equation}
\begin{equation}
\rho^{+}=\rho^{(0)} + \rho^{+(1)}\left({i\over 2}\lambda\right) + o(i\lambda).
\end{equation}
Unrelated to the external measuring field, the zero-order term is the result from the mean-field approach. By contrast, the 
concerned fluctuation of the mean electron occupation number induced by the Coulomb interaction is mainly described by the 
first-order term, and hence connected with the counting field. Actually, the two zero-order self-consistent equations hold the 
same form:
\begin{equation}\label{rho0exp}
\rho^{(0)} = \int {{dE}\over{2\pi}} {{\Gamma_{L} f_{L}+\Gamma_{R} f_{R}}\over{{\cal D}^{(0)}}},
\end{equation}
where ${\cal D}^{(0)}= (E-\epsilon_{0}^{(0)})^{2} + \Gamma^{2}$ and $\epsilon_{0}^{(0)}=E_{0}+U\rho^{(0)}$. Obviously, 
$\rho^{(0)}$ is totally determined by the Eq.(\ref{rho0exp}) , and will be assumed to be known in the solutions of $\rho^{-(1)}$ 
and $\rho^{+(1)}$.
Then, let us turn to the first-order self-consistent equations, and for the purpose of symmetric forms, they are constructed as:
\bn \label{solverho1}
\rho^{-(1)}+\rho^{+(1)} &=& ({\cal M}_{1} + {\cal M}_{2}) (\rho^{-(1)}-\rho^{+(1)}) \cr
 && + {\cal M}_{3}(\rho^{-(1)}+\rho^{+(1)}) +{\cal M}_{4},
\en
\bn \label{solverho2}
\rho^{-(1)}-\rho^{+(1)} &=& (-{\cal M}_{1} + {\cal M}_{5}) (\rho^{-(1)}+\rho^{+(1)}) \cr
 && + {\cal M}_{3}(\rho^{-(1)}-\rho^{+(1)}) +{\cal M}_{6},
\en
where the integrals in the equations are
\begin{subequations}
\begin{equation}\label{intpara1}
{\cal M}_{1} = \int {dE\over{2\pi i}} {U \over{{\cal D}^{(0)}}},
\end{equation}
\begin{equation}
{\cal M}_{2} = \int {dE\over{2\pi i}} {2U{\cal A}^{2}\over{{{\cal D}^{(0)}}^{2}}},
\end{equation}
\begin{equation}
{\cal M}_{3} = \int {dE\over{2\pi}} {2U{\cal A}(E-\epsilon_{0})\over{{{\cal D}^{(0)}}^{2}}},
\end{equation}
\begin{equation}
{\cal M}_{4} = \int {dE\over{2\pi}} {2{\cal A}{\cal B}\over{{{\cal D}^{(0)}}^{2}}},
\end{equation}
\begin{equation}
{\cal M}_{5} = \int {dE\over{2\pi i}} {2U(E-\epsilon_{0})^{2}\over{{{\cal D}^{(0)}}^{2}}},
\end{equation}
\begin{equation}\label{intpara2}
{\cal M}_{6} = \int {dE\over{2\pi i}} {2(E-\epsilon_{0}){\cal B}\over{{{\cal D}^{(0)}}^{2}}},
\end{equation}
\end{subequations}
with
\begin{subequations}
\begin{equation}
{\cal A} = \Gamma_{L}(f_{L}-{1\over{2}}) + \Gamma_{R}(f_{R}-{1\over{2}}),
\end{equation}
\begin{equation}
{\cal B} = \Gamma_{L} \Gamma_{R} (f_{R}-f_{L}).
\end{equation}
\end{subequations}
Solving the self-consistent Eq.(\ref{solverho1}) and Eq.(\ref{solverho2}), we find the solutions for $\rho^{-(1)}$ and 
$\rho^{+(1)}$, which account for the fluctuation of electron occupation number and contribute to the correction of the shot 
noise.

Until now, we have obtained the zero and first order solutions of GFs in theory if substituting the numerical calculations of 
$\rho^{(0)}$, $\rho^{-(1)}$ and $\rho^{+(1)}$ into $\hat{G}_{d\sigma}$. Then, it is natural to derive the formulae of the current 
$I$ and the zero-frequency shot noise $S$. According to Eq.(16), Eq.(17) and Eqs.(\ref{sigmasqd}), we arrive at the explicit 
expression of Eq.(\ref{adiabatic2})($\bar{\lambda}_{\eta}=\lambda_{\eta -}-\lambda_{\eta +}$):
\bn
 {\partial {\cal U}(\lambda_{-},\lambda_{+}) \over {\partial \lambda_{L -}}} &=& - \Gamma_{L} \Gamma_{R}\int {dE \over{2 \pi}} 
{1\over{{\cal D}}} \left [ e^{i\bar{\lambda}_{L}/2} f_{L} (1-f_{R}) \right. \cr
 && \left. - e^{-i\bar{\lambda}_{L}/2} f_{R} (1-f_{L}) \right ].
\en
Thus, the current can be calculated as
\bq \label{Iexpsqd}
I = {1 \over{\cal T}} {{\partial \ln{\chi(\lambda)}} \over{\partial(i\lambda_{L}/2)}} {\bigg |}_{\lambda=0}
  = 2\int {{d E}\over{2\pi}} T(E) (f_{L}-f_{R}),
\eq
with the transmission coefficient $T(E)$ defined as ($\Gamma = (\Gamma_{L} + \Gamma_{R})/2$):
\begin{equation}
T(E) = {{\Gamma_{L} \Gamma_{R}} \over{ (E-E_{0}-U\rho^{(0)})^{2} + \Gamma^{2}}}.
\end{equation}
And the zero-frequency shot noise can be evaluated as
\begin{equation}
S = 2{1 \over{\cal T}} {{\partial^{2} \ln{\chi(\lambda)}} \over{\partial(i\lambda_{L}/2)^{2}}} {\bigg |}_{\lambda=0} = S_{0} 
+S_{c}.
\end{equation}
 Here, we divide the zero-frequency shot noise $S$ into two separated parts, i.e. the mean-field current fluctuation $S_{0}$ and 
the correction term $S_{c}$, which is due to the intradot Coulomb interaction. Their explicit expressions are as follows:
\begin{widetext}
\begin{equation} \label{S0expsqd}
S_{0} =4\int {{dE}\over{2\pi}} \left\{ T(E)[f_{L}(1-f_{R})+f_{R}(1-f_{L})]-T^2(E)(f_{L}-f_{R})^{2} \right\},
\end{equation}
and
\begin{equation}\label{Scexp}
S_{c} = {4\over{\Gamma_{L}\Gamma_{R}}}\int {{dE}\over{2\pi}}  T^2(E) (f_{L}-f_{R}) U \left \{ (E-\epsilon_{0}) 
(\rho^{-(1)}+\rho^{+(1)})
  + i [\Gamma_{L}(f_{L}-{1\over{2}})+\Gamma_{R}(f_{R}-{1\over{2}})] (\rho^{-(1)}-\rho^{+(1)}) \right \}.
\end{equation}
\end{widetext}

About twenty years ago, S. Hershfield \cite{Hershfield} has adopted the Feynman diagram expansion technique to obtain the 
concerned expression of $S_{c}$ by some response and correlation functions. In order to compare our derivation of $S_{c}$ with 
this result, we begin with the definitions of the current-density response function $\chi_{jn}$, the density-density response 
function $\chi_{nn}$, the density-density correlation function $S_{nn}$ and the current-density correlation function $S_{jn}$:
\begin{subequations}
\begin{equation}
\chi_{jn}=\int dt \left\{ -i\theta(t) \langle[j(t),\rho(0)]\rangle \right\},
\end{equation}
\begin{equation}
\chi_{nn}=\int dt \left\{-i\theta(t) \langle[\rho(t),\rho(0)]\rangle \right\},
\end{equation}
\begin{equation}
S_{nn}=\int dt \left[\langle \rho(t) \rho(0) \rangle - \langle \rho \rangle^{2} \right],
\end{equation}
\begin{equation}
S_{jn}=\int dt \left[\langle j(t) \rho(0) \rangle+\langle \rho(t) j(0) \rangle - 2\langle j \rangle \langle \rho \rangle \right],
\end{equation}
\end{subequations}
where $j(t)$ indicates the current operator. One can define the effective distribution function of the quantum dot $f_{eff} = 
(\Gamma_{L}f_{L}+\Gamma_{R}f_{R})/(\Gamma_{L}+\Gamma_{R})$ and carry out the calculations as:
\begin{subequations}
\begin{equation}
\chi_{jn} = - \int {dE\over{2\pi}} {2\Gamma_{L}\Gamma_{R}\over{{{\cal D}^{(0)}}^{2}}} (f_{L}-f_{R})(E-\epsilon_{0}),
\end{equation}
\begin{equation}
\chi_{nn} =  4\Gamma \int {dE\over{2\pi}} {{E-\epsilon_{0}}\over{{{\cal D}^{(0)}}^{2}}}f_{eff},
\end{equation}
\begin{equation}
S_{nn} = 4\Gamma^{2} \int {dE\over{2\pi}} {f_{eff}(1-f_{eff})\over{{{\cal D}^{(0)}}^{2}}},
\end{equation}
\begin{equation}
S_{jn} = 2\Gamma_{L}\Gamma_{R}\Gamma \int {dE\over{2\pi}} {{1-2f_{eff}}\over{{{\cal D}^{(0)}}^{2}}} (f_{L}-f_{R}).
\end{equation}
\end{subequations}
Now it is convenient to have the integrals in Eqs.(\ref{intpara1}) $-$ (\ref{intpara2}) expressed in terms of the above response 
and correction functions, and hence $\rho^{-(1)}$ and $\rho^{+(1)}$ in terms of $\chi_{jn}$, $\chi_{nn}$, $S_{jn}$ and $S_{nn}$:
\bq \label{rho1ex1}
\rho^{-(1)}+\rho^{+(1)} = {S_{jn}\over{1-U\chi_{nn}}} + {2U\chi_{jn}S_{nn}\over{(1-U\chi_{nn})^{2}}}
\eq
\bq \label{rho1ex2}
\rho^{-(1)}-\rho^{+(1)} = {-i\chi_{jn}\over{1-U\chi_{nn}}}.
\eq
Substituting the Eq.(\ref{rho1ex1}) and (\ref{rho1ex2}) into Eq.(\ref{Scexp}) yields
\begin{equation}
\begin{array}{rrr}
S_{c} &=& 4 \left[ \chi_{jn} {{U}\over{1-U\chi_{nn}}} S_{nn} {{U}\over{1-U\chi_{nn}}} \chi_{jn} \right.  \\
 && + \left. S_{jn} {{U}\over{1-U\chi_{nn}}} \chi_{jn} \right] .
\end{array}
\end{equation}

Comparing this $S_{c}$ expression with the previous result of S. Hershfield, we believe the constant factor 2 before the second 
term on the right hand side of the equation in his result is not necessary. From physical point of view, obviously, this vertex
correction part of the shot noise is totally related to a set of response and correction functions of the electron density and 
the current. On the other hand, from technical point of view, the correction term corresponds to the vertex correction to the 
nonvanishing connection part in the conventional Feynman diagram expansion technique\cite{Ding}. Until now, we have obtain all 
the important equations we are interested in for the case of single quantum dot system.

\subsection{Results and Discussion}

For each set of chosen parameters, from the self-consistent equations (\ref{rho0exp}) $-$ (\ref{solverho2}), we obtain 
$\rho^{(0)}$, $\rho^{-(1)}$ and $\rho^{+(1)}$, which are then substituted in Eqs.(\ref{Iexpsqd}), (\ref{S0expsqd}) and 
(\ref{Scexp}) to compute the mean-field result of zero-frequency shot noise and its correction part. In the calculations, we set 
the Fermi energy zero, $E_{F}=0$ in equilibrium situation, and apply the bias voltage symmetrically, $\mu_{L,R} = \pm {1\over 2} 
V$. Without loss of generality, we assume $E_{0}=0$, $U=2\Gamma_{0}$ and $T=0.1\Gamma_{0}$ in the following discussion concerning 
the symmetric case and nonsymmetric case, where $\Gamma_{0}$ is the energy unit during numerical calculations through out this 
paper

For the symmetric case, Fig.1(a) plots $\rho^{(0)}$ and $|\rho^{(1)}|$ versus the external bias voltage (note 
$\rho^{-(1)}={\rho^{+(1)}}^{\ast}$). Obviously, the magnitude of $|\rho^{(1)}|$ is one order smaller than $\rho^{(0)}$ and the 
peak of $|\rho^{(1)}|$ corresponds to the large increasing rate of $\rho^{(0)}$ with respect to bias voltage. The first feature 
indicates small $S_{c}$ in comparison with $S_{0}$, because according to the fomula of zero-frequency shot noise, a small 
fluctuation of electron occupation number relative to $\rho^{(0)}$ exerts a small disturbance to the mean-field result $S_{0}$.

\begin{figure}[htb]
\includegraphics[height=4.3cm,width=8.7cm]{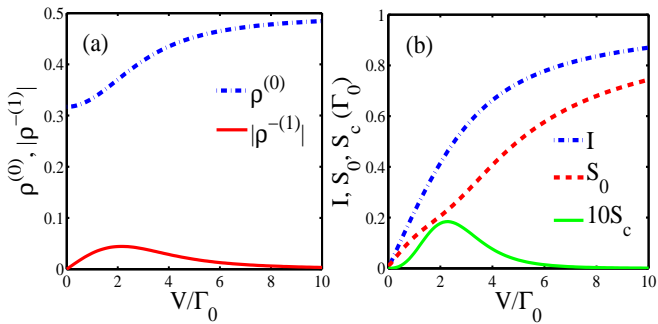}
\caption{(Color online) (a) Zero-order and first-order of electron occupation numbers versus bias voltage. (b) Current $I$, 
noninteracting noise $S_{0}$ and 10 times correction term of shot noise $10 S_{c}$ versus bias voltage. Parameters: 
$\Gamma_{L}=\Gamma_{R}=\Gamma_{0}$.}
\end{figure}

The second feature about the peak of $S_{c}$ can be explained by the resonant tunneling effect. Actually, the Hartree 
approximation gives the electron energy level $\epsilon_{0} = E_{0}+U\rho \simeq 0.8 \Gamma_{0}$ and hence the resonant tunneling 
bias voltage $1.6\Gamma_{0}$, if we use $\rho^{(0)}\sim 0.4$ to roughly estimate $\epsilon_{0}$ at the voltage point where 
$S_{c}$ reaches peak. It is an acceptable prediction about the tunneling voltage compared with the approximate result 
$2\Gamma_{0}$ indicated in Fig.1(a). In Fig.1(b), $S_{c}$ is nearly one order of magnitude smaller than $S_{0}$, which is 
consistent with the ratio of $|\rho^{(1)}|$ to $\rho^{(0)}$. In other word, the correction part of shot noise stemming from the 
fluctuation of electron occupation number is usually not prominent enough to increase the Fano factor largely enough and even 
greater than 1 (Fano factor $\gamma$ is defined as $\gamma= S/(2I)$).

For nonsymmetric cases, we define the nonsymmetric factor $\gamma_{ns}=\Gamma_{L} / \Gamma_{R}$ to describe the nonsymmetric 
degree. As illuminated in Fig.2, we plot the mean-field result term and the correction term of zero-frequency shot noise for 
$\gamma_{ns} = 3$ and $\gamma_{ns}={1\over 3}$ respectively. In both cases, the magnitude of $S_{c}$ is one order smaller than 
$S_{0}$, as in the symmetric case.
However, for nonsymmetric case we find that the correction term from the intradot Coulomb interaction contributes positively to 
the whole shot noise for $\gamma_{ns}={1\over 3}$, while for $\gamma_{ns} = 3$ the contribution turns out to be negative.
To interpret the sign of $S_{c}$, we would like to consider two extreme cases in the following discussions, , i.e. $\gamma_{ns} 
\gg 1$ and $\gamma_{ns} \ll 1$. In the case of $\gamma_{ns} \gg 1$, electrons flush into the dot from the left electrode but have 
little chance to be pumped out into the right lead, which results in the saturation of electron occupation inside the dot and a 
high potential ``wall'' (Coulomb blockade) in the dot. This strongly strengthened Coulomb blockade immediately reduces some 
possibilities of dynamic processes. For example, because of the strong Coulomb repulsion in the dot, the electrons are not as 
free to hop from the right lead back to the dot as they do if there are no Coulomb interaction in the dot,  which actually 
decreases the dynamic fluctuation of current. That is to say the interplay of Coulomb repulsion and many enough electrons pilling 
in the dot gives rise to the negative sign of $S_{c}$.

\begin{figure}[htb]
\includegraphics[height=4.3cm,width=8cm]{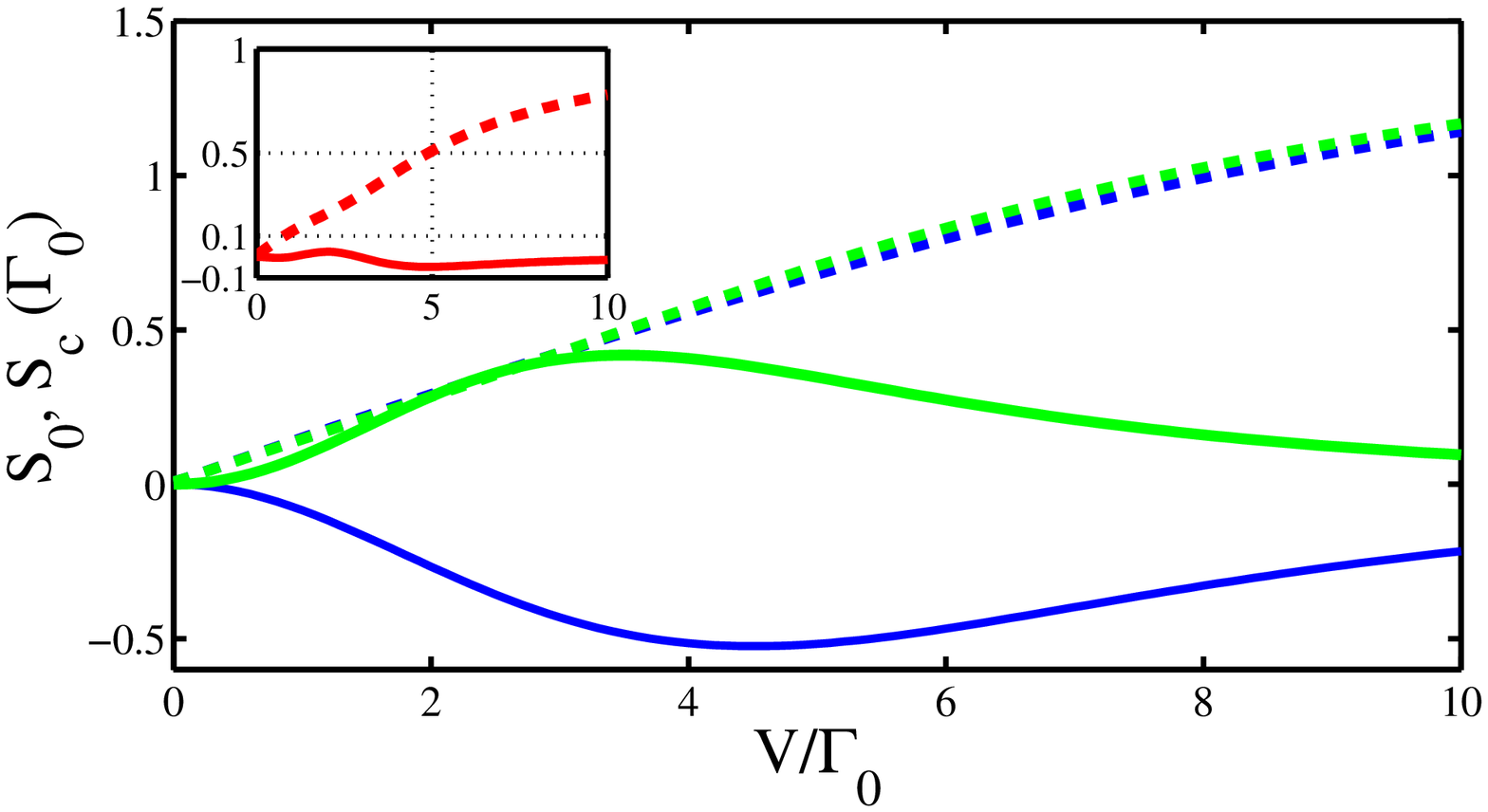}
\caption{(Color online) Noninteracting noise $S_{0}$ (dash) and 10 times correction term of shot noise $10 S_{c}$ (solid) versus 
bias voltage with $E_{0}=0$, $U=2\Gamma_{0}$ and $T=0.1\Gamma_{0}$ for $\gamma_{ns} = 3$ (blue), $\gamma_{ns} = {1 \over 3}$ 
(green). Inset: $S_{0}$ and $10 S_{c}$ are plotted for $\gamma_{ns} = 1.15$ (red).}
\end{figure}

On the contrary, similar analyses for the case of $\gamma_{ns} \ll 1$ show that Coulomb interaction leads to the increase of shot 
noise if few electrons occupy the dot. Thus, we speculate that the sign of $S_{c}$ is related to electron occupation numbers, 
which is associated with $\gamma_{ns}$, and the absolute contribution of $S_{c}$ may goes to nearly zero if a proper value of 
$\gamma_{ns}$ is chosen. For instance, the inset in Fig.2 illustrates the case of $\gamma_{ns} = 1.15$: the maximum absolute 
value of $10 S_{c}$ becomes one order of magnitude smaller, being 0.048.

\section{COUPLED QUANTUM DOTS}

\subsection{Model and Approximation}

We employ the two-impurity Anderson model to describe the system of coupled quantum dots (CQD), i.e. $H = H_{L} + H_{R} + H_{D} + 
H_{T}$, in which
\begin{equation}
H_{\eta}=\sum_{\eta k} \epsilon_{\eta k} c_{\eta k}^{\dagger} c_{\eta k},
\end{equation}
\begin{equation}\label{HdinCQD}
H_{D}= \epsilon_1 c_1^{\dagger} c_1 +\epsilon_2 c_2^{\dagger} c_2 +\Omega (c_1^{\dagger}c_2 + c_2^{\dagger} c_1) + U 
c_1^{\dagger} c_1 c_2^{\dagger} c_2,
\end{equation}
\begin{equation}
H_{T}= \sum_{k} \left[ V_{L}e^{i\lambda_{L}(t)/2}c_1^{\dagger}c_{Lk} + V_{R}e^{i\lambda_{R}(t)/2}c_{Rk}^{\dagger}c_2 + H.c. 
\right] .
\end{equation}
$H_{D}$ models the coupled quantum dots with $\epsilon_{1}$($\epsilon_{2}$) denoting the energy level of the quantum dot 1(2) and 
$\Omega$ indicating the magnitude of electron tunneling between the two dots, where we employ the single energy level 
approximation. As we just want to focus our attention on the effect of interdot Coulomb interaction $U$ in this case, we assume 
the Coulomb interaction in either single quantum dot goes to infinity and thus the spin index is omitted. $H_{T}$ denotes the 
tunneling between the reservoirs and two QDs, in which the parameter $V_{\eta}$ ($\eta=L,R$) is the corresponding tunneling 
magnitude between the dot and the electrode, and the measuring field ($\lambda_{\eta}$) is introduced in a symmetric way. For the 
convenience of further discussion, let $\lambda_{\eta -}=-\lambda_{\eta +}=\lambda_{\eta}$ and the symmetric configuration of 
external counting field requires that $\lambda_{L}=-\lambda_{R}=\lambda$.

In this case, we adopt the conventional HF approximation, which was suggested to be more reasonable than Hubbard-I approximation 
as long as the applied bias voltage is not relatively large\cite{Pedersen}, i.e.
\bn
c_{1}^{\dagger}c_{1}c_{2}^{\dagger}c_{2} &\simeq& \langle c_{1}^{\dagger}c_{1}\rangle c_{2}^{\dagger}c_{2} + c_{1}^{\dagger}c_{1} 
\langle c_{2}^{\dagger}c_{2} \rangle \cr && - \langle c_{1}^{\dagger}c_{2} \rangle c_{2}^{\dagger}c_{1} - \langle 
c_{2}^{\dagger}c_{1} \rangle c_{1}^{\dagger}c_{2}.\label{HCQDex}
\en
Then we introduce the effective energy levels $\epsilon_{\alpha r}=\epsilon_{\alpha}+U\rho_{\beta \beta}$ and the effective 
coupling parameter $\Omega_{\alpha \beta} = \Omega-U\rho_{\alpha \beta}$ ($\alpha \neq \beta$) with $\rho_{\alpha \beta}$ being 
defined as $\rho_{\alpha \beta}= \langle c_{\alpha}^{\dagger} c_{\beta} \rangle$ (${\alpha, \beta}={1,2}$). Thus, we can put 
Eq.(\ref{HdinCQD}) in a concise form:
\begin{equation}
H_{D} = \epsilon_{1r} c_1^{\dagger} c_1 + \epsilon_{2r} c_2^{\dagger} c_2 + \Omega_{12} c_1^{\dagger} c_2 + \Omega_{21} 
c_2^{\dagger} c_1.\label{hamitoninacqd}
\end{equation}
We still want to accentuate that depending on whether the time argument is on the forward path or the backward path on Keldysh 
contour, $\rho_{\alpha\beta}$ meets the obvious relation $\rho_{\alpha \beta}^{-} \neq \rho_{\alpha \beta}^{+}$, which results in 
$\epsilon_{\alpha r}^{-} \neq \epsilon_{\alpha r}^{+}$ and $\Omega_{\alpha \beta}^{-} \neq \Omega_{\alpha \beta}^{+}$.

\subsection{Theoretical Formulation}

As we can see from the single dot case, the to-be-solved expressions of the adiabatic potential, which play a central role in 
deriving the final results, counts on the solutions of a set of Green functions of the central region. So we employ EOM technique 
to solve these Green functions of the CQD, which are defined as
\bq
G_{\alpha\beta}(t,t' )= - i \langle \textit{T}_{C} c_{\alpha}(t) c_{\beta}^{\dagger}(t') \rangle_{\lambda}.
\eq
We introduce Green function of CQD in energy domain in the matrix form as
\begin{equation}
\hat{G}=
\left(
\begin{array}{cccc}
G_{11}^{--} & G_{11}^{-+} & G_{12}^{--} & G_{12}^{-+} \\
G_{11}^{+-} & G_{11}^{++} & G_{12}^{+-} & G_{12}^{++} \\
G_{21}^{--} & G_{21}^{-+} & G_{22}^{--} & G_{22}^{-+} \\
G_{21}^{+-} & G_{21}^{++} & G_{22}^{+-} & G_{22}^{++}
\end{array}
\right).
\end{equation}
Then some trivial calculations lead to the Dyson equation in the matrix form:
\begin{equation}\label{Dysonequation}
\hat{g}^{-1} \hat{G} = \hat{I} + \hat{\Sigma} \hat{G},
\end{equation}
where $\hat{g}$ denotes the uncoupled QD Green function matrix and $\hat{\Sigma}$ the self-energy matrix. All the details are 
illustrated clearly in the Appendix B. However, the Dyson equation is not enough to obtain the Green functions, $\hat{G}$, 
because the self-consistent parameters $\rho_{11}$, $\rho_{12}$, $\rho_{21}$ and $\rho_{22}$ in Eq.(\ref{Dysonequation}), which 
can be evaluated by the following self-consistent equations:
\begin{equation}\label{rho1cqdexp1}
\rho_{\alpha\beta}^{-} = \int {{dE}\over{2 \pi i}} G_{\alpha\beta}^{--}(E) e^{i E 0^{+}},
\end{equation}
\begin{equation}\label{rho1cqdexp2}
\rho_{\alpha\beta}^{+} = \int {{dE}\over{2 \pi i}} G_{\alpha\beta}^{++}(E) e^{-i E 0^{+}}.
\end{equation}
By introducing a projection operator $\hat{P}$ and integral operators $\hat{{\cal I}}_{-}$ and $\hat{{\cal I}}_{+}$, which are 
defined as $\hat{P} G_{\alpha\beta}^{\pm\pm} = G_{\alpha\beta}^{\pm\pm}$, $\hat{P} G_{\alpha\beta}^{\pm\mp} = 0$, $\hat{{\cal 
I}}_{-} = \int {{dE}\over{2 \pi i}} e^{i E 0^{+}}$ and $\hat{{\cal I}}_{+} = \int {{dE}\over{2 \pi i}} e^{-i E 0^{+}}$, we can 
rewrite the eight self-consistent equations concisely in a matrix form,
\begin{equation}\label{selfconsiscqd}
\hat{\rho} = \hat{\cal P} \hat{G},
\end{equation}
where the density matrix $\hat{\rho}$ is defined as
\begin{equation}
\hat{\rho}=
\left(
\begin{array}{cccc}
\rho_{11}^{-} & 0 & \rho_{21}^{-} & 0 \\
0 & \rho_{11}^{+} & 0 & \rho_{21}^{+} \\
\rho_{12}^{-} & 0 & \rho_{22}^{-} & 0 \\
0 & \rho_{12}^{+} & 0 & \rho_{22}^{+}
\end{array}
\right),
\end{equation}
and the operator matrix $\hat{\cal P}$ is defined as
\begin{equation}
{\cal P}=
\left(
\begin{array}{cccc}
\hat{{\cal I}}_{-} \hat{P}\\
& \hat{{\cal I}}_{+} \hat{P} &  & \text{{\huge{0}}}\\
& & \hat{{\cal I}}_{-} \hat{P} \\
& \text{{\huge{0}}} & & \hat{{\cal I}}_{+} \hat{P}
\end{array}
\right).
\end{equation}

Because only the current and shot noise are considered in this work, we only need to expand $\hat{\rho}$ to the zero-order and 
the first-order with respect to the counting field parameters $\lambda_{L}$ and $\lambda_{R}$:\cite{Dong3}
\begin{equation}
\hat{\rho} = \hat{\rho}^{(0)} + \sum_{\eta} \hat{\rho}_{\eta}^{(1)} \left({i\over2}\lambda_{\eta}\right) + 
o(\lambda_{L},\lambda_{R}),
\end{equation}
where the zero-order term $\hat{\rho}^{(0)}$ gives
\begin{equation}
\hat{\rho}^{(0)}=
\left(
\begin{array}{cccc}
\rho_{11}^{(0)} & 0 & \rho_{21}^{(0)} & 0 \\
0 & \rho_{11}^{(0)} & 0 & \rho_{21}^{(0)} \\
\rho_{12}^{(0)} & 0 & \rho_{22}^{(0)} & 0 \\
0 & \rho_{12}^{(0)} & 0 & \rho_{22}^{(0)}
\end{array}
\right),
\end{equation}
and the first-order term $\hat{\rho}_{\eta}^{(1)}$ gives
\begin{equation}
\hat{\rho}_{\eta}^{(1)}=
\left(
\begin{array}{cccc}
\rho_{11,\eta}^{-(1)} & 0 & \rho_{21,\eta}^{-(1)} & 0 \\
0 & \rho_{11,\eta}^{+(1)} & 0 & \rho_{21,\eta}^{+(1)} \\
\rho_{12,\eta}^{-(1)} & 0 & \rho_{22,\eta}^{-(1)} & 0 \\
0 & \rho_{12,\eta}^{+(1)} & 0 & \rho_{22,\eta}^{+(1)}
\end{array}
\right).
\end{equation}
Correspondingly, we expand ${\hat{g}}^{-1}$, $\hat{G}$ and $\hat{\Sigma}$ to the first-order with respect to $\lambda_{L}$ and 
$\lambda_{R}$:
\begin{equation}\label{gexpand}
{\hat{g}}^{-1} = {\hat{g}}^{-1(0)} + \sum_{\eta} {\hat{g}}^{-1(1)}_{\eta} \left({i\over 2}\lambda_{\eta}\right) + 
o(\lambda_{L},\lambda_{R}),
\end{equation}
\begin{equation}\label{Gexpand}
\hat{G} = \hat{G}^{(0)} +\sum_{\eta} \hat{G}_{\eta}^{(1)} \left({i\over 2}\lambda_{\eta}\right) + o(\lambda_{L},\lambda_{R}),
\end{equation}
\begin{equation}\label{Sigmaexpand}
\hat{\Sigma} = \hat{\Sigma}^{(0)} + \sum_{\eta} \hat{\Sigma}_{\eta}^{(1)} \left({i\over 2}\lambda_{\eta}\right) + 
o(\lambda_{L},\lambda_{R}),
\end{equation}
where the specific elements in the above matrixes are elaborated in Appendix B. Substituting 
Eqs.(\ref{gexpand})-(\ref{Sigmaexpand}) into Eq.(\ref{Dysonequation}), one obtains the formal solutions of the zero-order and the 
first-order terms of $\hat{G}$ in terms of $\hat{g}$ and $\hat{\Sigma}$:
\begin{equation}\label{G0DYSON}
\hat{G}^{(0)} = ({\hat{g}}^{-1(0)} - \hat{\Sigma}^{(0)})^{-1},
\end{equation}
\begin{equation}\label{G1LDYSON}
\hat{G}_{L}^{(1)} = \hat{G}^{(0)} (\hat{\Sigma}_{L}^{(1)} - {\hat{g}}^{-1(1)}_{L}) \hat{G}^{(0)},
\end{equation}
\begin{equation}\label{G1RDYSON}
\hat{G}_{R}^{(1)} = \hat{G}^{(0)} (\hat{\Sigma}_{R}^{(1)} - {\hat{g}}^{-1(1)}_{R}) \hat{G}^{(0)}.
\end{equation}
Note that Eq.(\ref{matrixg-1qcd}) tells the expression of ${\hat{g}}^{-1(0)}$ contains the elements of $\hat{\rho}^{(0)}$ and 
that of ${\hat{g}}^{-1(1)}_{\eta}$ contains the elements of $\hat{\rho}^{(0)}$ and $\hat{\rho}_{\eta}^{(1)}$. Inserting the 
expansions of $\hat{\rho}$  and $\hat{G}$ into Eq.(\ref{selfconsiscqd}), one immediately arrive at:
\begin{equation}\label{G0selfconsistent}
\hat{\rho}^{(0)} = \hat{{\cal P}} \hat{G}^{(0)},
\end{equation}
\begin{equation}\label{G1Lselfconsistent}
\hat{\rho}_{L}^{(1)} = \hat{{\cal P}} \hat{G}_{L}^{(1)},
\end{equation}
\begin{equation}
\hat{\rho}_{R}^{(1)} = \hat{{\cal P}} \hat{G}_{R}^{(1)}.
\end{equation}
\begin{widetext}
Now we are at a proper position to have $\hat{\rho}^{0}$ and $\hat{\rho}_{\eta}^{(1)}$ truly solved. Eq.(\ref{G0DYSON}) tells 
that all the elements of $\hat{G}^{(0)}$ are regarded as the function of the matrix $\hat{\rho}^{(0)}$, on the other hand 
$\hat{\rho}^{(0)}$ is determined by $\hat{G}^{(0)}$ as Eq.(\ref{G0selfconsistent}) indicates. Thus, combining the two matrix 
equations, we give the following equations to evaluate $\hat{\rho}^{0}$ self-consistently:
\begin{small}
\begin{subequations}
\begin{equation} \label{rho0expqcd11}
\rho_{11}^{(0)}= \int {{dE}\over{2\pi}} {1\over{{\Delta}^{(0)}}} \left\{ 
\Gamma_{L}f_{L}\left[(E-\epsilon_{2}-U\rho_{11}^{(0)})^{2}+{1\over{4}}\Gamma_{R}^{2}\right] + 
\Gamma_{R}f_{R}(\Omega-U\rho_{12}^{(0)})(\Omega-U\rho_{21}^{(0)}) \right\},
\end{equation}
\begin{equation}
\rho_{22}^{(0)}= \int {{dE}\over{2\pi}} {1\over{{\Delta}^{(0)}}} \left\{ 
\Gamma_{R}f_{R}\left[(E-\epsilon_{1}-U\rho_{22}^{(0)})^{2}+{1\over{4}}\Gamma_{R}^{2}\right] + 
\Gamma_{L}f_{L}(\Omega-U\rho_{12}^{(0)})(\Omega-U\rho_{21}^{(0)}) \right\},
\end{equation}
\begin{equation}
\rho_{21}^{(0)}= \int {{dE}\over{2\pi i}} {1\over{{\Delta}^{(0)}}} (\Omega-U\rho_{21}^{(0)}) \left\{ 
i\Gamma_{L}f_{L}(E-\epsilon_{2}-U\rho_{11}^{(0)}) + i\Gamma_{R}f_{R}(E-\epsilon_{1}-U\rho_{22}^{(0)}) - 
{1\over{2}}\Gamma_{L}\Gamma_{R}(f_{L}-f_{R})  \right\},
\end{equation}
\begin{equation}
\rho_{12}^{(0)}= \int {{dE}\over{2\pi i}} {1\over{{\Delta}^{(0)}}} (\Omega-U\rho_{12}^{(0)}) \left\{ 
i\Gamma_{L}f_{L}(E-\epsilon_{2}-U\rho_{11}^{(0)}) + i\Gamma_{R}f_{R}(E-\epsilon_{1}-U\rho_{22}^{(0)}) + 
{1\over{2}}\Gamma_{L}\Gamma_{R}(f_{L}-f_{R})  \right\}.
\end{equation}
\end{subequations}
\end{small}
Note that these zero-order electron occupation numbers hold the following relations: $\rho_{11}^{(0)\ast}= \rho_{11}^{(0)}$, 
$\rho_{22}^{(0)\ast}=\rho_{22}^{(0)}$ and $\rho_{12}^{(0)\ast}= \rho_{21}^{(0)\ast}$. With the solved $\hat{\rho}^{(0)}$ and 
$\hat{G}^{(0)}$, we turn to the solution of $\hat{\rho}_{L}^{(1)}$. Plugging Eq.(\ref{G1LDYSON}) into 
Eq.(\ref{G1Lselfconsistent}) yields the following linear algebraic equations:
\begin{small}
\begin{equation} \label{rho1matrixexpqcd}
\hat{\rho}_{L}^{(1)}
= \hat{{\cal P}} \hat{G}^{(0)}
\left(
\begin{array}{cccc}
0 & -i\Gamma_{L}f_{L} \\
-i\Gamma_{L}(1-f_{L}) & 0 & & \text{{\huge{0}}} \\
 & & & \\
 & \text{{\huge{0}}} &  & \text{{\huge{0}}}
\end{array}
\right)
\hat{G}^{(0)} - U \hat{{\cal P}} \hat{G}^{(0)}
\left(
\begin{array}{cccc}
-\rho_{22,L}^{-(1)} & 0 & \rho_{21,L}^{-(1)} & 0 \\
0 & \rho_{22,L}^{+(1)} & 0 & -\rho_{21,L}^{+(1)} \\
\rho_{12,L}^{-(1)} & 0 & -\rho_{11,L}^{-(1)} & 0 \\
0 & -\rho_{12,L}^{+(1)} & 0 & \rho_{11,L}^{+(1)}
\end{array}
\right)
\hat{G}^{(0)}.
\end{equation}
\end{small}
Note that similar linear algebraic equations for $\hat{\rho}_{R}^{(1)}$ can be reached.

All the preparations above pave for the discussion of the interested physics quantity, whose expressions will be derived in the 
rest of this section. The adiabatic potential is always the starting point according to the technique of FCS. According to 
nonequilibrium Feynman-Hellmann theorem, we have
\bn \label{adiabaticcqd}
{\partial {\cal U}(\lambda_{-},\lambda_{+}) \over {\partial \lambda_{L-}}} &=& \Big\langle {\partial H_{T}(t)\over{\partial
\lambda_{L-}}} \Big\rangle_\lambda = {i\over{2}} \sum_{k} \langle V_{L} e^{i\lambda_{L-}/2} c_{1}^{\dagger} c_{Lk} - V_{L} 
e^{-i\lambda_{L-}/2} c_{Lk}^{\dagger} c_{1} \rangle_{\lambda} \cr
&=& {1\over{2}} \sum_{k} \left[ V_{L} e^{i\lambda_{L-}/2} G_{Lk,1}^{--}(t,t^{+}) - V_{L} e^{-i\lambda_{L-}/2} 
G_{1,Lk}^{--}(t,t^{+}) \right].
\en
Applying the Keldysh disentanglement to the terms $G_{Lk,1}^{--}(t,t')$ and $G_{1,Lk}^{--}(t,t')$ in the integrand of 
Eq.(\ref{adiabaticcqd}), we have
\begin{equation}
{\partial {\cal U}(\lambda_{-},\lambda_{+}) \over {\partial \lambda_{L-}}} = {1\over{2}} \int d t_{1} \left[ 
\Sigma_{L}^{--}(t,t_{1}) G_{11}^{--}(t_{1},t) - \Sigma_{L}^{-+}(t,t_{1}) G_{11}^{+-}(t_{1},t) - G_{11}^{--}(t,t_{1}) 
\Sigma_{L}^{--}(t_{1},t) + G_{11}^{-+}(t,t_{1}) \Sigma_{L}^{+-}(t_{1},t) \right],
\end{equation}
where the self-energy is defined as $\Sigma_{\eta}^{\mu\nu} = \sum_{k} V_{\eta}^{2} e^{i(\lambda_{\eta\mu}-\lambda_{\eta\nu})/2} 
g_{\eta k}^{\mu\nu}$. Performing the Fourier transform and substituting the specific expressions of the corresponding self-energy 
and Green functions (Appendix B), one obtain
\begin{equation}
{\partial {\cal U}(\lambda_{-},\lambda_{+}) \over {\partial \lambda_{L-}}} = {1\over{2}} \Gamma_{L}\Gamma_{R} \int {{d 
E}\over{2\pi}} {1\over{\Delta}} \left\{ (\Omega-U\rho_{12}^{+})(\Omega-U\rho_{21}^{-})f_{R}(1-f_{L})e^{-i\tilde{\lambda}/2} - 
(\Omega-U\rho_{12}^{-})(\Omega-U\rho_{21}^{+})f_{L}(1-f_{R})e^{i\tilde{\lambda}/2} \right\},
\end{equation}
with $\tilde{\lambda} = \bar{\lambda}_{L} - \bar{\lambda}_{R}$ ($\bar{\lambda}_{\eta}=\lambda_{\eta-}-\lambda_{\eta+}$) and 
$\Delta$ being equal to the determinant of $({\hat{g}}^{-1}-\hat{\Sigma})$. Based on the identity within the adiabatic method 
${{\partial \ln\chi}\over{\partial(\lambda_{L}/2)}} = -i {\cal T} {{\partial {\cal U}}\over{\partial(\lambda_{L}/2)}}$, the 
current is given as
\begin{equation}
I = {1\over{{\cal T}}} {{\partial \ln\chi}\over{\partial (i \lambda_{L}/2)}}{\bigg |}_{\lambda=0} = \int {{d E}\over{2\pi}} 
T_{0}(E) (f_{L}-f_{R}),
\end{equation}
where the zero-order transmission coefficient $T_{0}(E)$ is defined as
\begin{equation}
T_{0}(E) = \Gamma_{L}\Gamma_{R}(\Omega-U\rho_{12}^{(0)}) (\Omega-U\rho_{21}^{(0)}) \left| 
(E-\epsilon_{1}-U\rho_{22}^{(0)}+{i\over{2}}\Gamma_{L}) (E-\epsilon_{2}-U\rho_{11}^{(0)}+{i\over{2}}\Gamma_{R}) - 
(\Omega-U\rho_{12}^{(0)})(\Omega-U\rho_{12}^{(0)})\right|^{-2}.
\end{equation}
When it comes to the expression of the zero-frequency shot noise, we first rewrite the formula for the adiabatic potential as
\begin{equation}
{\partial {\cal U}\over{\partial(\lambda_{L}/2)}} = \int {{dE}\over{2\pi}} { { \left[ 
T_{\lambda}^{(1)}(E)f_{R}(1-f_{L})e^{-i\tilde{\lambda}/2} - T_{\lambda}^{(2)}(E)f_{L}(1-f_{R})e^{i\tilde{\lambda}/2} \right]} 
\over {1 + \left[ T_{\lambda}^{(1)}(E)f_{R}(1-f_{L})(e^{-i\tilde{\lambda}/2}-1) + 
T_{\lambda}^{(2)}(E)f_{L}(1-f_{R})(e^{i\tilde{\lambda}/2}-1) \right] } },
\end{equation}
where we define the transmission coefficients $T_{\lambda}^{(1)}(E) = [\Gamma_{L}\Gamma_{R}(\Omega-U\rho_{12}^{+}) 
(\Omega-U\rho_{21}^{-})]/\Delta_{i}$ and $T_{\lambda}^{(2)}(E) = [\Gamma_{L}\Gamma_{R}(\Omega-U\rho_{12}^{-}) 
(\Omega-U\rho_{21}^{+})]/\Delta_{i}$ (refer to Eq.(\ref{deltaiexp}) in Appendix B for $\Delta_{i}$), which hold the relation 
$T_{\lambda}^{(1)}(E)|_{\lambda=0} = T_{\lambda}^{(2)}(E)|_{\lambda=0} = T_{0}(E)$. In this way, we make it technically natural 
to separate the shot noise into two parts, i.e.
\begin{equation}
S = 2{1\over{{\cal T}}} {{\partial \ln^{2} \chi}\over{\partial (i \lambda_{L}/2)^{2}}}{\bigg |}_{\lambda=0} = S^{(0)} + S^{(c)},
\end{equation}
where $S^{(0)}$ and $S^{(c)}$ are respectively the mean-field result and the correction part from the interdot Coulomb 
interaction. Actually, $S^{(0)}$ holds the same form as the single QD case, and is derived as
\begin{equation}
S^{(0)} = 2\int {dE\over{2\pi}} \left\{ T_{0}(E)[f_{R}(1-f_{L})+f_{L}(1-f_{R})] - T_{0}^{2}(E)(f_{L}-f_{R})^{2} \right\}.
\end{equation}
On the contrary, the explicit expression of $S^{(c)}$ is different from that of the single QD:
\bq
S^{(c)} = 2\int {{d E}\over{2\pi}} \left[ {{\partial T_{\lambda}^{(2)}(E)}\over{\partial (i\lambda_{L}/2)}}{\bigg |}_{\lambda=0} 
f_{L}(1-f_{R}) - {{\partial T_{\lambda}^{(1)}(E)}\over{\partial (i\lambda_{L}/2)}}{\bigg |}_{\lambda=0} f_{R}(1-f_{L}) \right].
\eq
Writing out the derivative of $T_{\lambda}(E)$ with respect to $(i\lambda_{L}/2)$, we find that $S^{(c)}$ can be split into two 
parts $S_{1}^{(c)}$ and $S_{2}^{(c)}$ corresponding respectively to the fluctuations of the electronic occupation numbers, 
$\rho_{\beta \beta}$, and coherence terms, $\rho_{\alpha \beta}$ ($\alpha \neq \beta$). We might as well define some auxiliary 
quantity to make their expressions more succinct: $\tilde{\epsilon}_{1}^{(0)}= E-\epsilon_{1} - U\rho_{22}^{(0)}$, 
$\tilde{\epsilon}_{2}^{(0)}= E-\epsilon_{2} - U\rho_{11}^{(0)}$, $\Omega_{12}^{(0)}= \Omega - U\rho_{12}^{(0)}$ and 
$\Omega_{21}^{(0)}= \Omega - U\rho_{21}^{(0)}$. Then we give the formula for $S_{1}^{(c)}$ as
\bn\label{Sc1qcd}
S_{1}^{(c)} &=& {2U\over{\Gamma_{L}\Gamma_{R}}} \int {{dE}\over{2\pi}} T_{0}^{2}(E) \left\{ \Omega_{12}^{(0)} \Omega_{21}^{(0)} 
(f_{L}-f_{R}) \left[
 {\cal K}_{1} (\rho_{22,L}^{-(1)}+\rho_{22,L}^{+(1)}) + {\cal K}_{2} (\rho_{11,L}^{-(1)}+\rho_{11,L}^{+(1)}) \right. \right. \cr
  && \left. \left. + {\cal K}_{3} (\rho_{22,L}^{-(1)}-\rho_{22,L}^{+(1)}) + {\cal K}_{4} (\rho_{11,L}^{-(1)}-\rho_{11,L}^{+(1)}) 
\right] \right\},
\en
with
\begin{subequations}
\bq
{\cal K}_{1} = \tilde{\epsilon}_{1}^{(0)}\left[(\tilde{\epsilon}_{2}^{(0)})^{2}+{1\over{4}}\Gamma_{R}^{2}\right] - 
\tilde{\epsilon}_{2}^{(0)} \Omega_{12}^{(0)}\Omega_{21}^{(0)},
\eq
\bq
{\cal K}_{2} = \tilde{\epsilon}_{2}^{(0)}\left[(\tilde{\epsilon}_{1}^{(0)})^{2}+{1\over{4}}\Gamma_{L}^{2}\right] - 
\tilde{\epsilon}_{1}^{(0)} \Omega_{12}^{(0)}\Omega_{21}^{(0)},
\eq
\bq
{\cal K}_{3} = i\Gamma_{L}\left(f_{L}-{1\over{2}}\right)\left[(\tilde{\epsilon}_{2}^{(0)})^{2}+{1\over{4}}\Gamma_{R}^{2}\right] - 
i\Gamma_{R}(f_{R}-{1\over{2}}) \Omega_{12}^{(0)}\Omega_{21}^{(0)},
\eq
\bq
{\cal K}_{4} = i\Gamma_{R}\left(f_{R}-{1\over{2}}\right)\left[(\tilde{\epsilon}_{1}^{(0)})^{2}+{1\over{4}}\Gamma_{L}^{2}\right] - 
i\Gamma_{L}(f_{L}-{1\over{2}}) \Omega_{12}^{(0)}\Omega_{21}^{(0)},
\eq
\end{subequations}
in which the fluctuations of electron occupation numbers, or the first-order terms in the expansion of $\rho_{11}$ and 
$\rho_{22}$ i.e. $\rho_{11,\eta}^{\mu (1)}$ and $\rho_{22,\eta}^{\mu (1)}$, are involved. For $S_{2}^{(c)}$, we have
\bn\label{Sc2qcd}
S_{2}^{(c)} &=& 2U \int {{dE}\over{2\pi}} \left\{ - {1\over{\Gamma_{L}\Gamma_{R}}} T_{0}^{2}(E) \Omega_{12}^{(0)} 
\Omega_{21}^{(0)} (f_{L}-f_{R}) \left[ {\cal L}_{1} (\rho_{12,L}^{-(1)}+\rho_{12,L}^{+(1)}) + {\cal L}_{2} 
(\rho_{21,L}^{-(1)}+\rho_{21,L}^{+(1)}) \right. \right. \cr && \left. + {\cal L}_{3} (\rho_{12,L}^{-(1)}-\rho_{12,L}^{+(1)}) + 
{\cal L}_{4} (\rho_{21,L}^{-(1)}-\rho_{21,L}^{+(1)}) \right] + \Omega_{21}^{(0)}T_{0}(E) \left[ \rho_{12,L}^{+(1)}f_{R}(1-f_{L}) 
- \rho_{12,L}^{-(1)}f_{L}(1-f_{R}) \right] \cr
&& \left. + \Omega_{12}^{(0)}T_{0}(E) \left[ \rho_{21,L}^{-(1)}f_{R}(1-f_{L}) - \rho_{21,L}^{+(1)}f_{L}(1-f_{R}) \right] 
\right\},
\en
with
\begin{subequations}
\bq
{\cal L}_{1}= \Omega_{21}^{(0)} (\tilde{\epsilon}_{1}^{(0)} \tilde{\epsilon}_{2}^{(0)} - \Omega_{12}^{(0)}\Omega_{21}^{(0)} 
-{1\over{4}}\Gamma_{L}\Gamma_{R})
\eq
\bq
{\cal L}_{2}= \Omega_{12}^{(0)} (\tilde{\epsilon}_{1}^{(0)} \tilde{\epsilon}_{2}^{(0)} - \Omega_{12}^{(0)}\Omega_{21}^{(0)} 
-{1\over{4}}\Gamma_{L}\Gamma_{R}),
\eq
\bq
{\cal L}_{3}= \Omega_{21}^{(0)} \left[ i\Gamma_{L}\left(f_{L}-{1\over{2}}\right)\tilde{\epsilon}_{2}^{(0)} + 
i\Gamma_{R}\left(f_{R}-{1\over{2}}\right)\tilde{\epsilon}_{1}^{(0)} - {1\over{2}} \Gamma_{L}\Gamma_{R}(f_{L}-f_{R}) \right],
\eq
\bq
{\cal L}_{4}= \Omega_{12}^{(0)} \left[ i\Gamma_{L}\left(f_{L}-{1\over{2}}\right)\tilde{\epsilon}_{2}^{(0)} + 
i\Gamma_{R}\left(f_{R}-{1\over{2}}\right)\tilde{\epsilon}_{1}^{(0)} + {1\over{2}} \Gamma_{L}\Gamma_{R}(f_{L}-f_{R}) \right],
\eq
\end{subequations}
which is related to the first-order terms in the expansions of $\rho_{12}$ and $\rho_{21}$, i.e. $\rho_{12,\eta}^{\mu (1)}$ and 
$\rho_{21,\eta}^{\mu (1)}$. Thus, according to which elements in the first-order terms of density matrixes 
($\hat{\rho}_{\eta}^{(1)}$) are involved in the expression of correction part of shot noise, we divide the formula of $S^{(c)}$ 
into $S_{1}^{(c)}$ and $S_{2}^{(c)}$. Considering the similarity between $S_{1}^{(c)}$ and $S^{(c)}$, which is also related to 
the fluctuation of electron occupation number, we can expect the absolute value of $S_{1}^{(c)}$ is much smaller than $S^{(0)}$. 
However, different from the single dot system, in the coupled dots system the new additional correction part of zero-frequency 
shot noise, $S_{2}^{(c)}$, stems from the coherence dynamic fluctuations ($\rho_{12,\eta}^{\mu (1)}$ and $\rho_{21,\eta}^{\mu 
(1)}$), which can play a dominating role in increasing shot noise and Fano factor ($\gamma$), and can even give rise to $\gamma 
>1$ under certain circumstances, as illustrated in the following section.

\begin{figure}[hbt]
\includegraphics[height=9.5cm,width=17cm]{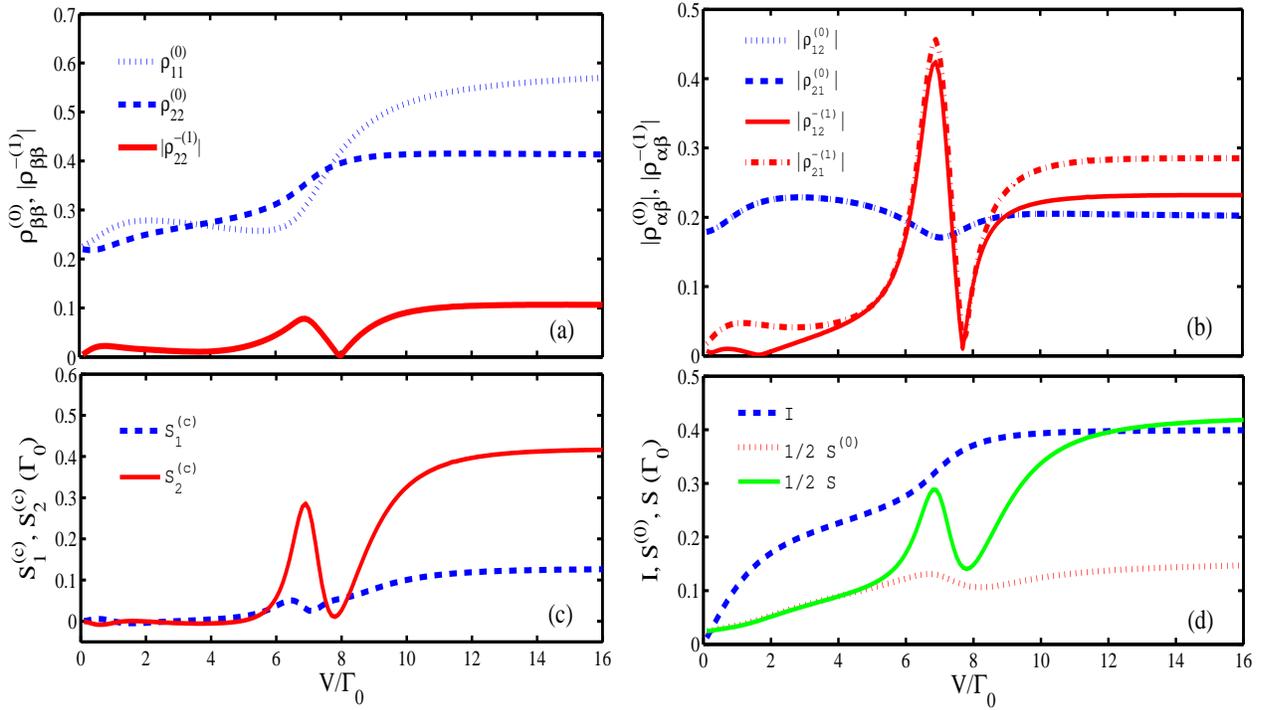}
\caption{(Color online) (a) Zero-order terms ($\rho^{(0)}_{\beta\beta}$) and first-order terms ($|\rho^{-(1)}_{\beta\beta}|$) of 
electron occupation numbers versus bias voltage; (b) Zero-order coherence terms ($|\rho^{(0)}_{\alpha\beta}|$) and first-order 
coherence terms ($|\rho^{-(1)}_{\alpha\beta}|$) versus bias voltage, where we let $\alpha \neq \beta$ here; (c) Two separate 
components $S_{1}^{(c)}$ and $S_{2}^{(c)}$ of the shot noise correction terms versus bias voltage; (d) Current $I$, 
noninteracting noise $S^{(0)}$ and total shot noise $S$ versus bias voltage. Other Parameters used in the calculations: 
$\epsilon_{1}=\epsilon_{2}=\Gamma_{0}$, $\Gamma_{L}=\Gamma_{R}=\Gamma_{0}$, $\Omega=\Gamma_{0}$, $U=4\Gamma_{0}$, and 
$T=0.1\Gamma_{0}$.}
\end{figure}
\end{widetext}

\subsection{Results and Discussion}

All the elements of $\hat{\rho}^{(0)}$ and $\hat{\rho}_{\eta}^{(1)}$ are computed from self-consistent 
Eqs.~(\ref{rho0expqcd11})--(\ref{rho1matrixexpqcd}) and then are used to evaluate the current and the zero-frequency shot noise. 
In the following discussions, the correction part of shot noise resulting from the coherence fluctuations are highlighted to find 
the parameter window for a large Fano factor $\gamma$, especially for $\gamma > 1$.

We consider a typical symmetric case with $\Gamma_{L}=\Gamma_{R}=\Gamma_{0}$ and $\epsilon_{1}=\epsilon_{2}=\Gamma_{0}$, and 
thereby plot the calculated results in Fig.~3. In Fig.~3(a), the electron occupation numbers $\rho_{11}^{(0)}$ and 
$\rho_{22}^{(0)}$ are plotted as functions of the bias voltage, together with the bias dependence of $|\rho_{22}^{-(1)}|$ that 
reflects the nonequilibrium density fluctuation (NDF) of $\rho_{22}$. In Fig.~3(b), we plot the coherence terms, 
($|\rho_{12}^{(0)}|$, $|\rho_{21}^{(0)}|$), and the nonequilibrium coherence fluctuations (NCFs), ($|\rho_{12}^{-(1)}|$, 
$|\rho_{21}^{-(1)}|$) as functions of the bias voltage. In Fig.~3(c), $S_{1}^{(c)}$ and $S_{2}^{(c)}$ versus the bias voltage are 
depicted. From these figures, we find three features in the symmetric case: (1) The ratio of $|\rho_{12}^{-(1)}|$ 
($|\rho_{21}^{-(1)}|$) to $\rho_{12}^{(0)}$ ($\rho_{21}^{(0)}$) is much larger than the ratio of $|\rho_{22}^{-(1)}|$ to 
$\rho_{22}^{(0)}$ at nearly whole regions of the bias voltages, except for the region of relatively small bias voltages; (2) The 
second correction term, $S_{2}^{(c)}$, causes the main contribution to the shot noise of a CQD, other than the first correction 
term, $S_{1}^{(c)}$. As far as the resulted formulae of $S_{1}^{(c)}$ and $S_{2}^{(c)}$, i.e., Eqs.~(\ref{Sc1qcd}) and 
(\ref{Sc2qcd}) in the above subsection, are concerned, this feature can be ascribed to the first feature from numerical 
calculation point of view. In consequence, we can thereby deduce that the NCF effect plays a more prominent role in determining 
the zero-frequency shot noise of a CQD system; (3) $|\rho_{22}^{-(1)}|$, $|\rho_{12}^{-(1)}|$, $|\rho_{21}^{-(1)}|$, 
$S_{2}^{(c)}$ and $S$ all exhibit the peak and the plateau structures respectively with increasing bias voltage. The peaks are 
located at the value of bias voltage where a large change rate of electron occupation number takes place, being around 
$V=7\Gamma_{0}$, which can be explained by one of the approximate eigenvalues $\epsilon_{\alpha r} + \Omega_{\alpha \beta}$. 
While the plateaus appear when the bias voltage becomes greater than
$V=2(\epsilon_{\alpha} + U)=10\Gamma_{0}$.
In addition, Eq.(\ref{Sc2qcd}) requires that $S_{2}^{(c)}$ and $|\rho_{\alpha \beta}^{-(1)}|$ share the two structures nearly at 
the same values of bias voltages.
To understand such connections among $|\rho_{22}^{-(1)}|$, $|\rho_{12}^{-(1)}|$ and $|\rho_{21}^{-(1)}|$, one needs
to notice the HF energy levels for the two dots are $\epsilon_{\alpha}+U\rho_{\beta \beta}$ ($\alpha \neq \beta$). That is to say 
that the energy level of the dot $\alpha$ becomes uncertain at nonequilibrium condition and fluctuates roughly between 
$\epsilon_{\alpha}+U(\rho_{\beta
\beta}^{(0)} \pm |\rho_{\beta \beta}^{-(1)}|)$ due to the NDF effect. Its average energy level is $\epsilon_{\alpha}+U\rho_{\beta 
\beta}^{(0)}$.
Since the gap between energy levels of two QDs directly affects the overlap between the two electronic wavefunctions of the two 
dots, the energy level fluctuations induced by NDFs naturally result in the NCFs, which accounts for the reason that the peak and 
plateau structures appear almost at the same values of bias voltages for $|\rho_{22}^{-(1)}|$, $|\rho_{12}^{-(1)}|$ and 
$|\rho_{21}^{-(1)}|$. Surprisingly, a weak super-Poissonian shot noise is found from Fig.~3(d) at the considerably large bias 
voltage region, $V>12\Gamma_0$, which is different from previous results based on the quantum rate equation 
(QRE).\cite{Kieblich1,Dong1,Dong2,Kieblich2} These earlier work predict no super-Poissonian noise for the symmetric CQD in the 
limit of infinite bias voltage at zero temperature. Notice that the QRE approach is only valid for sequential tunneling and 
thereby fails to account for quantum coherence between two dots. Meanwhile, notice that the enhancement of shot noise is mainly 
coming from the contribution of $S_2^{(c)}$, i.e., the NCFs, in the present studies. We can thereby argue that the coherent 
exchange interaction effect is somehow accounted in certain degree in the present calculations of current fluctuations at the 
level of HF approximation, and causes the consequence of the enhanced noise.

\begin{figure*}
\includegraphics[height=9.5cm,width=17cm]{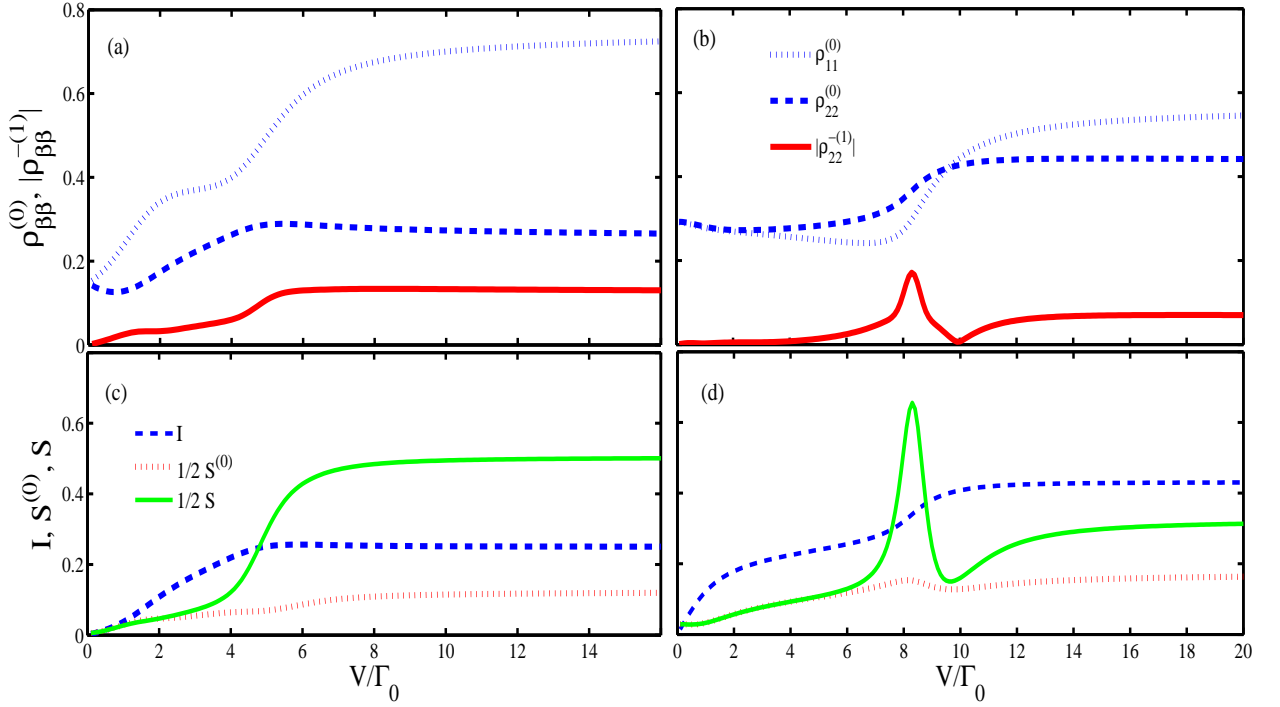}
\caption{(Color online) Zero-order terms ($\rho^{(0)}_{\beta\beta}$) and first-order terms ($|\rho^{-(1)}_{\beta\beta}|$) of 
electron occupation numbers versus bias voltage in (a, b), and current $I$, noninteracting noise $S^{(0)}$ and total shot noise 
$S$ versus bias voltage in (c, d). The case of $\Omega=0.5\Gamma_{0}$ and $U=2\Gamma_{0}$ is plotted in (a, c) and the case of 
$\Omega=1.25\Gamma_{0}$ and $U=5\Gamma_{0}$ in (b, d). Other Parameters are the same as those in Fig.~3.}
\end{figure*}

In order to find out the optimal parameter values for the super-Poissonian noise ($\gamma >1$), it is necessary in the following 
to investigate the two structures more deeply.
On the premise of keeping the ratio $U/\Omega = 4$ unchanged, we first study another two symmetric systems, i.e. (i) $U=2$, 
$\Omega=0.5$ and (ii) $U=5$, $\Omega=1.25$, and depict their results in Figs.~4(a,c) and Figs.4(b,d), respectively. For the sake 
of convenience, we temporarily name the previous system of $U=4$ and $\Omega=1$ as the system (iii).
For the systems (ii) and (iii), the peak of $|\rho_{22}^{-(1)}|$ corresponds to the large slopes or gradients of 
$\rho_{22}^{(0)}$ with respect to the bias voltage, while the peak disappears in the system (i) although there is still a large 
slope for $\rho_{22}^{(0)}$ as shown in Fig.~4(a). To figure out why the NCFs show different behaviors in these systems, we 
notice that the interdot coupling parameter $\Omega$ of the systems (ii) and (iii) is much larger than the system (i). It is 
understandable that a larger ratio of $\Omega$ to $\Gamma_{\eta}$ means electrons being much easier to hop from dot 1 to dot 2 
than electrons tunneling between the left lead and dot 1 or between the right lead and dot 2. As a result, a larger $\Omega$ 
relative to $\Gamma_{\eta}$ can narrow down the difference between $\rho_{11}^{(0)}$ and $\rho_{22}^{(0)}$ and hence the gap 
between two renormalized energy levels. Since a smaller energy gap implies stronger electronic wavefunction overlap between two 
dots, we can therefore ascribe the peak structure at the finite bias voltage to the stronger coherence and NCFs in the systems 
(ii) and (iii). Furthermore, we find that the system (ii) even shows a super-Poissonian noise due to the strong peak in the 
noise-voltage characteristics at a small bias region (noticing that this system has the strongest hopping $\Omega=1.25 \Gamma_0$ 
in our calculations). This new noise enhancement at small bias region is also different from previous prediction by Aghassi, in 
which a thermally excited multi-electron state is involved in sequential tunneling process and thus crucial for an enhanced noisy 
current.\cite{Aghassi}

With regard to the plateau structure, one can observe that the plateau is always accompanied by prominent difference between 
$\rho_{11}^{(0)}$ and $\rho_{22}^{(0)}$ at the large bias voltage region. As we indicated above, bigger density difference will 
cause more stronger NCF and inevitably cause more remarkable enhancement of the shot noise. This fact interprets the occurrence 
of weak super-Poissonian noise in the system (iii) at the large bias voltage region as shown in Fig.~3(d). Meanwhile, the system 
(i) with the smallest hopping $\Omega=0.5\Gamma_0$ will have the most biggest density difference in the present study and thereby 
exhibits the most obvious super-Poissonian noise at the large bias voltage region as display in Fig.~4(c).
Until now, we demonstrate that on the premise of setting $U/\Omega$ constant, a larger inter-dot hopping $\Omega$ induces a 
remarkable peak at small bias voltages, while a smaller hopping $\Omega$ leads to a prominent plateau structure at large bias 
voltages.

After these discussions with keeping $U/\Omega$ constant, we now consider different relative values of $U$ to $\Omega$. We plot 
Fano factor versus the bias voltage for different values of $U$ with $\Omega = 1$ in Fig.~5(a) and different values of $\Omega$ 
with $U=4$ in Fig.~5(b). From these results, we can argue that the Fano factor $\gamma$ is, in a certain degree, positively 
associated with the relative ratio $U/\Omega$. In fact, we notice that the HF approximation used in this paper defines an 
effective electron hopping parameter, $\Omega[1 - (U/\Omega) \rho_{\alpha \beta}]$, as indicated in the Hamiltonian 
Eq.~(\ref{hamitoninacqd}). As a result, it seems reasonable that we can regard the ratio $U/\Omega$ as a magnifying factor of the 
interdot coherence ($\rho_{12}^{(0)}$ and $\rho_{21}^{(0)}$) and hence of the NCFs ($\rho_{12}^{(1)}$ and $\rho_{21}^{(1)}$). 
Since $S_{2}^{(c)}$ is positively correlated to the NCFs, $\rho_{12}^{(1)}$ and $\rho_{21}^{(1)}$, the larger ratio $U/\Omega$ 
will naturally cause the larger $S_{2}^{(c)}$. It is worth noticing that this argument agrees qualitatively well with the 
rate-equation investigation by G. Kie{\ss}lich {\it et.al.} on the noise enhancement for the CQD system, in which the authors 
ascribed the noise enhancement to the interplay of Coulomb interaction and coherence.\cite{Kieblich1} Finally, we show the Fano 
factor as functions of the bias voltage and the undressed hopping parameter $\Omega$ with a fixed $U=4\Gamma_0$ in Fig.~6.

\begin{figure}
\includegraphics[height=4cm,width=8.5cm]{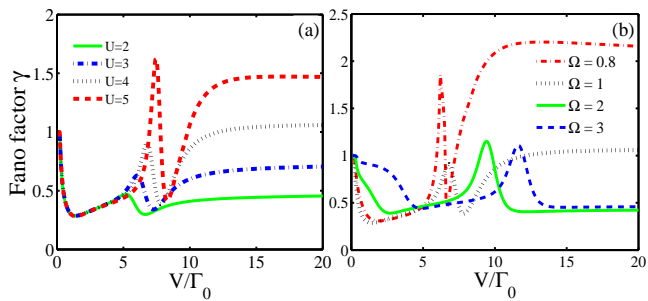}
\caption{(Color online) Fano factor versus bias voltage for a fixed $\Omega=1$ in (a) and a fixed $U=4$ in (b), respectively. The 
other parameters are the same as those in Fig.~3.}
\end{figure}

\begin{figure}
\includegraphics[height=6cm,width=8cm]{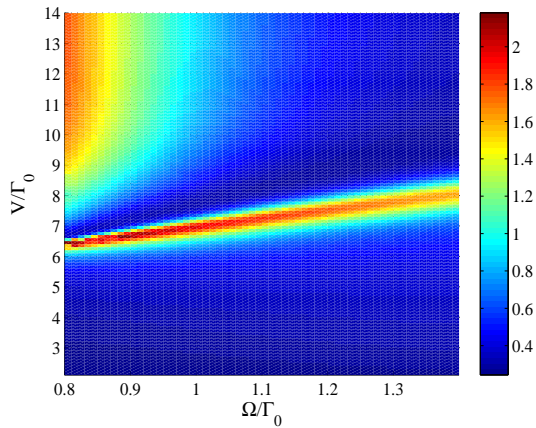}
\caption{(Color online) Fano factor versus bias voltage $V$ and interdot coupling parameter $\Omega$ with the following 
parameters: $\epsilon_{1}=\epsilon_{2}=\Gamma_{0}$, $U=4\Gamma_{0}$, and $T=0.1\Gamma_{0}$.}
\end{figure}

\section{CONCLUSIONS}

In this paper, we have investigated the zero-frequency shot noise of electronic tunneling through a SQD and a CQD focusing our 
attention on the effects of the intradot or interdot Coulomb interactions, respectively, within the theoretical framework of full 
counting statistics. Employing NGF techniques, we have derived the analytical expressions for the current and shot noise by 
self-consistent solutions within the Hartree approximation in a SQD system and the HF approximation in a CQD system.

For the SQD system, the obtained explicit expression for the correction term of zero-frequency shot noise, $S_{c}$, caused by the 
intradot Coulomb interaction, has been found to be in agreement with the earlier result by Hershfield using the Feynman diagram 
expansion method. Our numerical calculations have shown that the contribution of the correlation term is nearly negligible in 
comparison with the noninteracting shot noise (at least one order of magnitude smaller in our calculations for chosen 
parameters). In addition, it is interestingly found that altering the relative values of $\Gamma_{L}$ to $\Gamma_{R}$ can change 
the sign of $S_{c}$.

The main objective of this paper is to examine the zero-frequency shot noise in a symmetric CQD by taking account of the joint 
effects of Coulomb interaction and coherent exchange between two dots  beyond the usual rate-equation investigations. For this 
purpose, we have adopted the HF approximation to deal with the interdot Coulomb interaction, which is proved to be a reliable 
method for electronic tunneling through a CQD as long as the applied bias voltage is not relatively large.\cite{Pedersen} We have 
found that the correction term of the zero-frequency shot noise of a CQD can be separated into two parts: one is coming from the 
contribution of the NDFs and the other one is originated from the contribution of the interdot NCFs. The numerical calculations 
have predicted that the interdot NCFs play a predominated role in determining the shot noise over the NDFs, and give rise to two 
characteristic structures, a peak and a plateau, in the behavior of bias-voltage-dependent shot noise, whose values may be 
enhanced to show super-Poissonian even for the symmetric CQD. It is worth pointing out that this result is new and different from 
those of rate-equation calculations, which
predicted either no super-Poissonian noise or super-Poissonian noise due to the thermal excited multi-electron state in the 
symmetric case. We hope that the present paper can stimulate further experimental investigation for shot noise in a CQD system.

\begin{acknowledgments}

This work was supported by Projects of the National Basic Research Program of China (973 Program) under Grant No.2011CB925603, 
and the National Science Foundation of China, Specialized Research Fund for the Doctoral Program of Higher Education (SRFDP) of 
China.
\end{acknowledgments}

\begin{appendix}

\section{Calculation of the contour-order GFs for the SQD}

In the appendix A, we employ the EOM technique to derive the contour-order GFs of the central region for the SQD system.

Considering the time evolution of $d_{\sigma}(t)$, one can easily obtain the equations of motion for $g_{d\sigma}^{--}(t,t')$ and 
$g_{d\sigma}^{++}(t,t')$ and then perform the Fourier transformation to give
\begin{equation}
\hat{g}_{d\sigma}^{-1}=
\left(
\begin{array}{cc}
E-\epsilon_{0}^{-} & 0 \\ 0 & -(E-\epsilon_{0}^{+})
\end{array}
\right).
\end{equation}
Next, one can proceed in the similar way to obtain the EOMs for GFs of the interacting quantum dot:
\begin{widetext}
\begin{equation}
 \left( i{\partial \over{\partial t}} - \epsilon_{0}^{\mu} \right) G_{d\sigma}^{\mu \nu}(t,t') = \delta_{C}(t-t') + \sum_{\eta k} 
V_{\eta k} e^{i\lambda_{\eta \mu}/2} G_{\eta k \sigma,d \sigma}^{\mu \nu}(t,t'),
\end{equation}
where the contour order Dirac delta function $\delta_{C}(t-t')$ is defined as ($C_{-}$ indicates the forward pathway and $C_{+}$ 
indicates the backward pathway on the Keldysh contour): $\delta_{C}(t-t') = \delta(t-t') (t,t' \in C_{-})$, $\delta_{C}(t-t') = - 
\delta(t-t')$ ($t,t' \in C_{+}$) and $\delta_{C}(t-t')=0$ ($t \in C_{-}, t' \in C_{+}$ or $t \in C_{+}, t' \in C_{-}$).
After Applying the Keldysh disentanglement, i.e.
\bq
G_{\eta k \sigma, d \sigma}^{\mu \nu}(t,t') = V_{\eta k} \int dt_{1} \left[ g_{\eta k\sigma}^{\mu -}(t,t_{1}) e^{-i\lambda_{\eta 
-}/2} G_{d\sigma}^{-\nu}(t_{1},t') - g_{\eta k\sigma}^{\mu +}(t,t_{1}) e^{-i \lambda_{\eta +}/2} G_{d\sigma}^{+\nu}(t_{1},t') 
\right],
\eq
and
\bq
G_{d \sigma ,\eta k \sigma}^{\mu \nu}(t,t') = V_{\eta k} \int dt_{1} \left[ G_{d\sigma}^{\mu -}(t,t_{1}) e^{i \lambda_{\eta -}/2} 
g_{\eta k\sigma}^{-\nu}(t_{1},t') - G_{d\sigma}^{\mu +}(t,t_{1}) e^{i \lambda_{\eta +}/2} g_{\eta k\sigma}^{+\nu}(t_{1},t') 
\right],
\eq
and the Fourier transformation, the above Green functions are reconstructed in a concise matrix form of Dyson equation 
$\hat{G}_{d\sigma} = \hat{g}_{d\sigma} + \hat{g}_{d\sigma} \hat{\Sigma} \hat{G}_{d \sigma}$, where $\hat{\Sigma}$ is the 
introduced self-energy matrix, expressed as
\begin{equation}
\hat{\Sigma}=
\left(
\begin{array}{cc}
\Sigma_{L\sigma}^{--} + \Sigma_{R\sigma}^{--} & -(\Sigma_{L\sigma}^{-+} + \Sigma_{R\sigma}^{-+}) \\ -(\Sigma_{L\sigma}^{+-} + 
\Sigma_{R\sigma}^{+-}) & \Sigma_{L\sigma}^{++} + \Sigma_{R\sigma}^{++}
\end{array}
\right).
\end{equation}

Here, $\Sigma_{\eta \sigma}^{\mu \nu}$ ($\eta= L,R$) is the self-energy. The explicit expressions for $\Sigma_{\eta \sigma}^{\mu 
\nu}$ are listed as follows:
\begin{equation} \label{sigmasqd}
\begin{array}{llll}
\Sigma_{L\sigma}^{--} = i \Gamma_{L} (f_{L}-{1 \over{2}}), & \Sigma_{L\sigma}^{-+} = i e^{i\bar{\lambda}_{L}/2}\Gamma_{L} f_{L}, 
& \Sigma_{L\sigma}^{+-} = -i \Gamma_{L} e^{-i\bar{\lambda}_{L}/2} (1-f_{L}), & \Sigma_{L\sigma}^{++} = i \Gamma_{L} (f_{L}-{1 
\over{2}}),\\
\Sigma_{R\sigma}^{--} = i \Gamma_{R} (f_{R}-{1 \over{2}}), & \Sigma_{R\sigma}^{-+} = i \Gamma_{R} f_{R}, & \Sigma_{R\sigma}^{+-} 
= -i \Gamma_{R} (1-f_{R}), & \Sigma_{R\sigma}^{++} = i \Gamma_{R} (f_{R}-{1 \over{2}}).
\end{array}
\end{equation}
Therefore, the corresponding GFs can be told from $\hat{G}_{d\sigma}=(\hat{g}_{d\sigma}^{-1}-\hat{\Sigma})^{-1}$, which gives 
Eq.(\ref{sqdGexpression}).

\section{Calculations of the contour-order GFs for the CQD and its expansion with respect to the counting fields}

In this appendix, we first apply the EOM technique to calculate the GFs for the CQD system, and then perform expansion of the 
resulted Dyson equation in to zero-order and first-order with respect to the counting fields.

We first bring in the matrix $\hat{g}$ and construct the equations of motion for all its elements into the matrix form in energy 
domain as
\begin{equation}\label{hatgex}
\hat{g}^{-1}
=
\left(
\begin{array}{cccc}
g_{11}^{--} & 0 & g_{12}^{--} & 0 \\
0 & g_{11}^{++} & 0 & g_{12}^{++} \\
g_{21}^{++} & 0 & g_{22}^{++} & 0 \\
0 & g_{21}^{++} & 0 & g_{22}^{++}
\end{array}
\right)^{-1}
=
\left(
\begin{array}{cccc}
E-\epsilon_{1r}^{-} & 0 & -\Omega_{12}^{-} & 0 \\
0 & -(E-\epsilon_{1r}^{+}) & 0 & \Omega_{12}^{+} \\
-\Omega_{21}^{-} & 0 & E-\epsilon_{2r}^{-} & 0 \\
0 & \Omega_{21}^{+} & 0 & -(E-\epsilon_{2r}^{+})
\end{array}
\right),
\end{equation}
Note that uncoupled QD Green functions with $t$ and $t'$ on the different pathways on the Keldysh contour vanish if only the 
final steady state is discussed. The dot Keldysh nonequilibrium Green functions meet the following equations of motion:
\begin{subequations}
\begin{equation}\label{EOMforCQD1}
\left( i{\partial\over{\partial t}}-\epsilon_{1r}^{\mu} \right) G_{11}^{\mu \nu}(t,t') = \delta_{C}(t-t') + 
\sum_{k}e^{i\lambda_{L\mu}/2}V_{L}G_{Lk,1}^{\mu \nu}(t,t')
  + \Omega_{12}^{\mu} G_{21}^{\mu \nu}(t,t'),
\end{equation}
\begin{equation}
\left(i{\partial\over{\partial t}}-\epsilon_{2r}^{\mu} \right) G_{22}^{\mu \nu}(t,t') = \delta_{C}(t-t') + 
\sum_{k}e^{i\lambda_{R\mu}/2}V_{R}G_{Rk,2}^{\mu \nu}(t,t') + \Omega_{21}^{\mu} G_{12}^{\mu \nu}(t,t'),
\end{equation}
\begin{equation}
\left(i{\partial\over{\partial t}}-\epsilon_{1r}^{\mu} \right) G_{12}^{\mu \nu}(t,t') =  
\sum_{k}e^{i\lambda_{L\mu}/2}V_{L}G_{Lk,2}^{\mu \nu}(t,t')+ \Omega_{12}^{\mu} G_{22}^{\mu \nu}(t,t'),
\end{equation}
\begin{equation}\label{EOMforCQD2}
\left(i{\partial\over{\partial t}}-\epsilon_{2r}^{\mu} \right) G_{21}^{\mu \nu}(t,t') =  
\sum_{k}e^{i\lambda_{R\mu}/2}V_{R}G_{Rk,1}^{\mu \nu}(t,t')+ \Omega_{21}^{\mu} G_{11}^{\mu \nu}(t,t'),
\end{equation}
\end{subequations}
where the mixed Green functions $G_{\eta k,1}(t,t')$ and $G_{1,\eta k}(t,t')$ are defined as
\begin{equation}
G_{\eta k,\alpha}(t,t')= -i \langle{ \textit{T}_{C}} c_{\eta k}(t) c_{\alpha}^{\dagger}(t') \rangle_{\lambda},
\text{{ }} \text{{ }} \text{{ }}
G_{\alpha,\eta k}(t,t')= -i \langle{ \textit{T}_{C}} c_{\alpha}(t) c_{\eta k}^{\dagger}(t') \rangle_{\lambda}.
\end{equation}
After applying Fourier transformation, Eqs.(\ref{EOMforCQD1})-(\ref{EOMforCQD2}) are constructed as the matrix form of Dyson 
equation, i.e. Eq.(\ref{Dysonequation}), where we introduce the self-energy matrix
\bq
\hat{\Sigma}
=
\left(
\begin{array}{cccc}
\Sigma_{L}^{--} & -\Sigma_{L}^{-+} & 0 & 0 \\
-\Sigma_{L}^{+-} & \Sigma_{L}^{++} & 0 & 0 \\
0 & 0 & \Sigma_{R}^{--} & -\Sigma_{R}^{-+} \\
0 & 0 & -\Sigma_{R}^{+-} & \Sigma_{R}^{++}
\end{array}
\right),
\eq
and all the elements of $\hat{\Sigma}$ are listed as follows:
\begin{equation}
\begin{array}{llll}
\Sigma_{L}^{--}=i\Gamma_{L}(f_{L}-{1\over 2}), & \Sigma_{L}^{-+}=ie^{i\bar{\lambda}_{L}/2}\Gamma_{L}f_{L}, &
\Sigma_{L}^{+-}=-ie^{-i\bar{\lambda}_{L}/2}\Gamma_{L}(1-f_{L}), & \Sigma_{L}^{++}=i\Gamma_{L}(f_{L}-{1\over 2}), \\
\Sigma_{R}^{--}=i\Gamma_{R}(f_{R}-{1\over 2}), & \Sigma_{R}^{-+}=ie^{i\bar{\lambda}_{R}/2}\Gamma_{R}f_{R}, &
\Sigma_{R}^{+-}=-ie^{-i\bar{\lambda}_{R}/2}\Gamma_{R}(1-f_{R}), & \Sigma_{R}^{++}=i\Gamma_{R}(f_{R}-{1\over 2}).
\end{array}
\end{equation}
Then the solution of the Dyson equation is found as $\hat{G}=({\hat{g}}^{-1}-\hat{\Sigma})^{-1}$. Let $\Delta$ equal to the 
determinant of $({\hat{g}}^{-1}-\hat{\Sigma})$, which requires being solved first before the solution of the Dyson equation. To 
have the expression of $\hat{G}$ brevity, we define some auxiliary quantities: $\tilde{\epsilon}_{1}^{-}= E-\epsilon_{1r}^{-}$, 
$\tilde{\epsilon}_{1}^{+}= -(E-\epsilon_{1r}^{+})$, $\tilde{\epsilon}_{2}^{-}= E-\epsilon_{2r}^{-}$, and 
$\tilde{\epsilon}_{2}^{+}= -(E-\epsilon_{2r}^{+})$.
Now, let's turn to the expression of $\Delta$ ($\tilde{\lambda}=\bar{\lambda}_{L}-\bar{\lambda}_{R}$):
\bn
\Delta &=& \left[(\tilde{\epsilon}_{1}^{-}-\Sigma_{L}^{--})(\tilde{\epsilon}_{1}^{+}-\Sigma_{L}^{++})- 
\Sigma_{L}^{-+}\Sigma_{L}^{+-}\right] \left[(\tilde{\epsilon}_{2}^{-}-\Sigma_{R}^{--})(\tilde{\epsilon}_{2}^{+}-\Sigma_{R}^{++})- 
\Sigma_{R}^{-+}\Sigma_{R}^{+-}\right] +(\Sigma_{L}^{-+}\Sigma_{R}^{+-}+\Omega_{12}^{-}\Omega_{21}^{+}) 
(\Sigma_{L}^{+-}\Sigma_{R}^{-+} \cr
&& +\Omega_{21}^{-}\Omega_{12}^{+}) - \Sigma_{L}^{-+}\Sigma_{L}^{+-}\Sigma_{R}^{-+}\Sigma_{R}^{+-}
-\Omega_{21}^{+}\Omega_{12}^{+}(\tilde{\epsilon}_{1}^{-}-\Sigma_{L}^{--})(\tilde{\epsilon}_{2}^{-}-\Sigma_{R}^{--}) - 
\Omega_{12}^{-}\Omega_{21}^{-}(\tilde{\epsilon}_{1}^{+}-\Sigma_{L}^{++})(\tilde{\epsilon}_{2}^{+}-\Sigma_{R}^{++}).
\en
We divide $\Delta$ into two separate parts, i.e $\Delta=\Delta_{i}+\Delta_{e}$. The term $\Delta_{e}$ explicitly contains the 
counting field parameter $\lambda_{\eta}$ and vanishes when setting $\lambda_{\eta}=0$:
\begin{equation}
\Delta_{e}= \Gamma_{L}\Gamma_{R}[(\Omega-U\rho_{21}^{-})(\Omega-U\rho_{12}^{+})f_{R}(1-f_{L})(e^{-i\tilde{\lambda}/2}-1) + 
(\Omega-U\rho_{21}^{+})(\Omega-U\rho_{12}^{-})f_{L}(1-f_{R})(e^{i\tilde{\lambda}/2}-1)].
\end{equation}
And the term $\Delta_{i}$, which implicitly contains the parameter $\lambda$ and is involved in the expression of 
$T_{\lambda}(E)$, gives the following formidable expression
\bn \label{deltaiexp}
\Delta_{i} &=& \left\{\left[ E-\epsilon_{1}-U\rho_{22}^{-}-i\Gamma_{L}(f_{L}-{1\over{2}})\right] 
\left[E-\epsilon_{1}-U\rho_{22}^{+}+i\Gamma_{L}(f_{L}-{1\over{2}})\right] + \Gamma_{L}^{2}f_{L}(1-f_{L})\right\} \cr
 && \times \left\{ \left[E-\epsilon_{2}-U\rho_{11}^{-}-i\Gamma_{R}(f_{R}-{1\over{2}})\right] 
\left[E-\epsilon_{2}-U\rho_{11}^{+}+i\Gamma_{R}(f_{R}-{1\over{2}})\right] + \Gamma_{R}^{2}f_{R}(1-f_{R})\right\} \cr
 && - (\Omega-U\rho_{12}^{+})(\Omega-U\rho_{21}^{+}) \left[ E-\epsilon_{1}-U\rho_{22}^{-}-i\Gamma_{L}(f_{L}-{1\over{2}}) \right] 
\left[ E-\epsilon_{2}-U\rho_{11}^{-}-i\Gamma_{R}(f_{R}-{1\over{2}}) \right] \cr
 && - (\Omega-U\rho_{12}^{-})(\Omega-U\rho_{21}^{-}) \left[ E-\epsilon_{1}-U\rho_{22}^{+}+i\Gamma_{L}(f_{L}-{1\over{2}}) \right] 
\left[ E-\epsilon_{2}-U\rho_{11}^{+}+i\Gamma_{R}(f_{R}-{1\over{2}}) \right] \cr
 && + \Gamma_{L}\Gamma_{R} \left[ (\Omega-U\rho_{21}^{-}) (\Omega-U\rho_{12}^{+}) f_{R}(1-f_{L}) + (\Omega-U\rho_{21}^{+}) 
(\Omega-U\rho_{12}^{-}) f_{L}(1-f_{R}) \right] \cr
 && + (\Omega-U\rho_{12}^{-})(\Omega-U\rho_{12}^{+})(\Omega-U\rho_{21}^{-})(\Omega-U\rho_{21}^{+}).
\en
This complicated expression illustrates the complexity of coupled dots system in theory. However, $\Delta_{i}$ can be reduced to 
a fairly simple result, when setting $\lambda=0$, which is the naturally requirement in deriving the current and shot noise 
formulas, as
\begin{equation}
\Delta^{(0)} \equiv \Delta_{i}(\lambda=0) = {\bigg |}(E-\epsilon_{1}-U\rho_{22}^{(0)}+{i\over{2}}\Gamma_{L}) 
(E-\epsilon_{2}-U\rho_{11}^{(0)}+{i\over{2}}\Gamma_{R}) - (\Omega-U\rho_{12}^{(0)})(\Omega-U\rho_{12}^{(0)}) {\bigg |}^{2},
\end{equation}
where $|...|$ stands for the plural mode. Now, it is natural to give the expression of $\hat{G}$ from the Dyson equation
\begin{equation} \label{Gdysonsolutionqdc}
\hat{G} = {1\over{\Delta}}
\left(
\begin{array}{cccc}
\tilde{\epsilon}_{1}^{-}-\Sigma_{L}^{--} & \Sigma_{L}^{-+} & -\Omega_{12}^{-} & 0 \\
\Sigma_{L}^{+-} & \tilde{\epsilon}_{1}^{+}-\Sigma_{L}^{++} & 0 & \Omega_{12}^{+} \\
-\Omega_{21}^{-} & 0 & \tilde{\epsilon}_{2}^{-}-\Sigma_{R}^{--} & \Sigma_{R}^{-+} \\
0 & \Omega_{21}^{+} & \Sigma_{R}^{+-} & \tilde{\epsilon}_{2}^{+}-\Sigma_{R}^{++} \\
\end{array}
\right)^{-1}
\end{equation}
By the way, in the derivation for the adiabatic potential, $G_{11}^{-+}$ and $G_{11}^{+-}$ are needed. So we give their explicit 
expressions here as
\begin{equation}
G_{11}^{+-}= - {1\over{\Delta}} 
[\Sigma_{L}^{+-}(\tilde{\epsilon}_{2}^{-}-\Sigma_{R}^{--})(\tilde{\epsilon}_{2}^{+}-\Sigma_{R}^{++}) - 
\Omega_{21}^{-}\Omega_{12}^{+}\Sigma_{R}^{+-} - \Sigma_{L}^{+-}\Sigma_{R}^{+-}\Sigma_{R}^{-+}],
\end{equation}
\begin{equation}
G_{11}^{-+}= - {1\over{\Delta}} 
[\Sigma_{L}^{-+}(\tilde{\epsilon}_{2}^{-}-\Sigma_{R}^{--})(\tilde{\epsilon}_{2}^{+}-\Sigma_{R}^{++}) - 
\Omega_{21}^{+}\Omega_{12}^{-}\Sigma_{R}^{-+} - \Sigma_{L}^{-+}\Sigma_{R}^{+-}\Sigma_{R}^{-+}].
\end{equation}

As illustrated before, the solution of $\hat{G}$ depends on the self-consistent parameters $\rho_{\alpha \beta}$, which have to 
be calculated by the self-consistent equations. We find that doing the expansion with respect to the parameter $\lambda_{L}$ and 
$\lambda_{R}$ for those well-defined matrix in the Dyson equation makes the solution feasible in practice. Therefore, we 
elaborate the expansion calculations here, especially specifying Eqs.(\ref{gexpand}) $-$ (\ref{Sigmaexpand}):
\bn
\hat{g}^{-1} &=&
\left(
\begin{array}{cccc}
E-\epsilon_{1}-U\rho_{22}^{(0)} & 0 & -\Omega+U\rho_{21}^{(0)} & 0 \\
0 & -(E-\epsilon_{1}-U\rho_{22}^{(0)}) & 0 & \Omega-U\rho_{21}^{(0)} \\
-\Omega+U\rho_{12}^{(0)} & 0 & E-\epsilon_{2}-U\rho_{11}^{(0)} & 0 \\
0 & \Omega-U\rho_{12}^{(0)} & 0 & -(E-\epsilon_{2}-U\rho_{11}^{(0)})
\end{array}
\right) \cr
 && + \sum_{\eta=L,R} \left({i\over 2}\lambda_{\eta}\right)
\left(
\begin{array}{cccc}
-U\rho_{22,\eta}^{-(1)} & 0 & U\rho_{21,\eta}^{-(1)} & 0 \\
0 & U\rho_{22,\eta}^{+(1)} & 0 & -U\rho_{21,\eta}^{+(1)} \\
U\rho_{12,\eta}^{-(1)} & 0 & -U\rho_{11,\eta}^{-(1)} & 0 \\
0 & -U\rho_{12,\eta}^{+(1)} & 0 & U\rho_{11,\eta}^{+(1)}
\end{array}
\right)
+ o(\lambda_{L},\lambda_{R}), \label{matrixg-1qcd}
\en
or being written in an abbreviating symbolic way as ${\hat{g}}^{-1} = {\hat{g}}^{-1(0)} + \sum_{\eta} \hat{g}^{-1(1)}_{\eta} 
\left({i\over 2}\lambda_{\eta}\right) + o(\lambda_{L},\lambda_{R})$,
\bn
 \hat{\Sigma} &=&
\left(
\begin{array}{cccc}
i\Gamma_{L}(f_{L}-{1\over{2}}) & -i\Gamma_{L}f_{L} & 0 & 0 \\
i\Gamma_{L}(1-f_{L}) & i\Gamma_{L}(f_{L}-{1\over{2}}) & 0 & 0 \\
0 & 0 & i\Gamma_{R}(f_{R}-{1\over{2}}) & -i\Gamma_{R}f_{R} \\
0 & 0 & i\Gamma_{R}(1-f_{R}) & i\Gamma_{R}(f_{R}-{1\over{2}})
\end{array}
\right)
+ \left({i\over 2}\lambda_{L}\right)
\left(
\begin{array}{cccc}
0 & -i\Gamma_{L}f_{L} \\
-i\Gamma_{L}(1-f_{L}) & 0 & & \text{{\huge{0}}} \\
 & \text{{\huge{0}}} & & \text{{\huge{0}}}\\
 & & & \\
\end{array}
\right)  \cr
 && + \left({i\over 2}\lambda_{R}\right)
\left(
\begin{array}{cccc}
\text{{\huge{0}}} & & \text{{\huge{0}}}\\
 & & & \\
\text{{\huge{0}}} & & 0 & -i\Gamma_{R}f_{R} \\
 & & -i\Gamma_{R}(1-f_{R}) & 0
\end{array}
\right)
+ o(\lambda_{L},\lambda_{R}),
\en
or symbolized as $\hat{\Sigma} = \hat{\Sigma}^{(0)} + \sum_{\eta} \hat{\Sigma}_{\eta}^{(1)} \left({i\over 2}\lambda_{L}\right) + 
o(\lambda_{L},\lambda_{R})$, and
\begin{equation}
\hat{G} =
\left(
\begin{array}{cccc}
G_{11}^{(0)} & G_{11}^{(0)} & G_{12}^{(0)} & G_{12}^{(0)} \\
G_{11}^{(0)} & G_{11}^{(0)} & G_{12}^{(0)} & G_{12}^{(0)} \\
G_{21}^{(0)} & G_{21}^{(0)} & G_{22}^{(0)} & G_{22}^{(0)} \\
G_{21}^{(0)} & G_{21}^{(0)} & G_{22}^{(0)} & G_{22}^{(0)}
\end{array}
\right)
+ \sum_{\eta=L,R} \left({i \over 2}\lambda_{\eta}\right)
\left(
\begin{array}{cccc}
G_{11,\eta}^{--(1)} & G_{11,\eta}^{-+(1)} & G_{12,\eta}^{--(1)} & G_{12,\eta}^{-+(1)} \\
G_{11,\eta}^{+-(1)} & G_{11,\eta}^{++(1)} & G_{12,\eta}^{+-(1)} & G_{12,\eta}^{++(1)} \\
G_{21,\eta}^{--(1)} & G_{21,\eta}^{-+(1)} & G_{22,\eta}^{--(1)} & G_{22,\eta}^{-+(1)} \\
G_{21,\eta}^{+-(1)} & G_{21,\eta}^{++(1)} & G_{22,\eta}^{+-(1)} & G_{22,\eta}^{++(1)}
\end{array}
\right)
+ o(\lambda_{L},\lambda_{R}),
\end{equation}
or symbolized as $\hat{G} = \hat{G}^{(0)} + \sum_{\eta} \hat{G}_{\eta}^{(1)} \left({i\over2}\lambda_{\eta}\right) + 
o(\lambda_{L},\lambda_{R})$.
Inserting all the expansions above in the Dyson equation (\ref{Dysonequation}), one obtains
\bq\label{dysoneqexpansioninqcd}
\left[{\hat{g}}^{-1(0)} + \sum_{\eta} \hat{g}^{-1(1)}_{\eta} \left({i\over 2}\lambda_{\eta}\right)\right]\left[\hat{G}^{(0)} + 
\sum_{\eta} \hat{G}_{\eta}^{(1)} \left({i\over 2}\lambda_{\eta}\right) \right] = \hat{I} + \left[\hat{\Sigma}^{(0)} + \sum_{\eta} 
\hat{\Sigma}_{\eta}^{(1)} \left({i\over 2}\lambda_{\eta}\right) \right] \left[\hat{G}^{(0)} + \sum_{\eta} \hat{G}_{\eta}^{(1)} 
\left({i\over 2}\lambda_{\eta}\right) \right].
\eq
Then if the second and higher orders are ignored, Eq.(\ref{dysoneqexpansioninqcd}) gives the zero-order and first-order Dyson 
equations, i.e. Eqs.(\ref{G0DYSON}) - (\ref{G1RDYSON}).
\end{widetext}

\end{appendix}

\end{document}